\documentclass[12pt]{article}

\usepackage[table]{xcolor}
\usepackage[normalem]{ulem}
\usepackage{booktabs}
\usepackage{longtable}
\usepackage{footmisc}

\usepackage{scicite}
\usepackage{ref_macros_sci}
\newcounter{firstbib}

\usepackage{graphicx}
\usepackage{setspace}

\usepackage{graphicx}
\usepackage{amsmath}
\usepackage{amssymb}
\usepackage{xspace}
\usepackage{multirow}

\usepackage{rotating}

\usepackage[final]{pdfpages}

\usepackage{times}

\newenvironment{sciabstract}{%
\begin{quote} \bf}
{\end{quote}}

\newcounter{lastnote}
\newenvironment{scilastnote}{%
\setcounter{lastnote}{\value{enumiv}}%
\addtocounter{lastnote}{+1}%
\begin{list}%
{\arabic{lastnote}.}
{\setlength{\leftmargin}{.22in}}
{\setlength{\labelsep}{.5em}}}
{\end{list}}

\title{A Remarkably Loud Quasi-Periodicity after a Star is Disrupted by a Massive Black Hole}

\author
{Dheeraj R. Pasham$^{1\ast}$, Ronald A. Remillard$^{1}$, P. Chris Fragile$^{2}$, Alessia \\
Franchini$^{3}$, Nicholas C. Stone$^{4}$, Giuseppe Lodato$^{3}$, Jeroen Homan$^{1}$, Deepto \\
Chakrabarty$^{1}$, Frederick K. Baganoff$^{1}$, James F. Steiner$^{1}$, Eric R. Coughlin$^{5}$, \\
Nishanth R. Pasham$^{6}$\\
\normalsize{$^{1}$MIT Kavli Institute for Astrophysics and Space Research, Cambridge, MA 02139}\\
\normalsize{$^{2}$Department of Physics and Astronomy, College of Charleston, Charleston, SC 29424}\\
\normalsize{$^{3}$Department of Physics, University of Milan, Dipartimento di Fisica, Via Celoria 16}\\
\normalsize{$^{4}$Department of Physics, Columbia University, New York, NY 10027}\\
\normalsize{$^{5}$Department of Physics, University of California, Berkeley, CA 94720-3411}\\
\normalsize{$^{6}$Allyo, Sunnyvale, CA}\\
\\
\normalsize{$^\ast$To whom correspondence should be addressed; E-mail:  dheeraj@space.mit.edu}
}

\date{}
\begin{document}

\baselineskip24pt
\maketitle 

\begin{sciabstract}

The immense tidal forces of massive black holes can rip apart stars that come too close to them. As the resulting stellar debris spirals inwards, it heats up and emits x-rays when near the black hole. Here, we report the discovery of an exceptionally stable 131-second x-ray quasi-periodicity from a black hole after it disrupted a star. Using a black hole mass indicated from host galaxy scaling relations implies that, (1) this periodicity originates from very close to the black hole's event horizon, and (2) the black hole is rapidly spinning. Our findings suggest that other disruption events with similar highly sensitive observations likely also exhibit quasi-periodicities that encode information about the fundamental properties of their black holes. 

\end{sciabstract}

Almost all massive galaxies are expected to harbor a massive black hole (MBH; masses $\gtrsim$10$^{4}$ solar masses, $M_{\odot}$) at their centers\cite{Richstone98}, yet most of them are inactive and do not produce any observable radiative output. However, it is predicted that roughly once every $\sim$10$^{4-5}$ years\cite{stonemetger_tde_rates,Holoien16,sjoert_tde_rates} a star will pass near enough to the BH to be disrupted by the BH's gravitational forces. Such episodes, known as tidal disruption events (TDEs)\cite{rees88}, trigger accretion onto quiescent BHs and provide a brief window of opportunity to measure the two fundamental properties that define astrophysical BHs: mass and spin. While there are empirical scaling laws to infer BH masses, for example, using host galaxy properties\cite{gultekin09}, the spins of massive BHs have been very difficult to constrain. This is because the effects of spin predicted by Einstein's theory of general relativity only emerge in the immediate vicinity of BHs, typically within a few gravitational radii\footnote{The gravitational radius, $R_{g}$ = $GM/c^{2}$, where $G$, $M$, and $c$ are the gravitational constant, BH mass, and the speed of light, respectively.}\cite{bardeen72}. Thus, BH spin measurements require studying radiation from the innermost regions of the accretion flow, where gravity is strong. Theoretical models for TDEs predict that shortly after the disruption, a fraction of the stellar debris settles into a hot thermal inner disk that emits primarily in the soft x-rays or extreme UV\cite{lodatorossi11}. Identifying such disk-dominated/soft-state x-ray bright TDEs can provide a new avenue to measure spins of numerous BHs lying dormant in external galaxies. However until now, x-ray data of known soft-state TDEs lacked the sensitivity to probe this strong-gravity regime in-detail. Here, we report on the TDE ASASSN-14li, which has been observed by all major x-ray telescopes and has yielded the most sensitive soft x-ray observations of a TDE to-date.

The spectacular transient event ASASSN-14li was discovered by the All-Sky Automated Survey for SuperNovae (ASASSN) on 22 November 2014\cite{Holoien16}. Soon after its discovery it became apparent that it exhibited nearly all properties of previously-known TDEs: a spatial position consistent with the host galaxy's center (within 160 parsecs\cite{Holoien16}), a luminosity declining in time with a power-law index of 5/3\cite{miller15} as expected from a TDE\cite{phinney89}, a blue optical spectrum with broad Hydrogen and Helium emission lines and a constant optical color unlike any ordinary supernova\cite{Holoien16}. In addition to the optical and UV emission, ASASSN-14li also produced an x-ray\cite{miller15} and a radio synchrotron flare\cite{alexander16,vanvelzen16}. Owing to these multi-wavelength properties ASASSN-14li has been dubbed the ``Rosetta stone'' for TDEs\cite{krolik16}.

\begin{figure}[ht]
\begin{center}
%\hspace{-0.35in}
\includegraphics[width=6.5in, height=5.85in, angle=0]{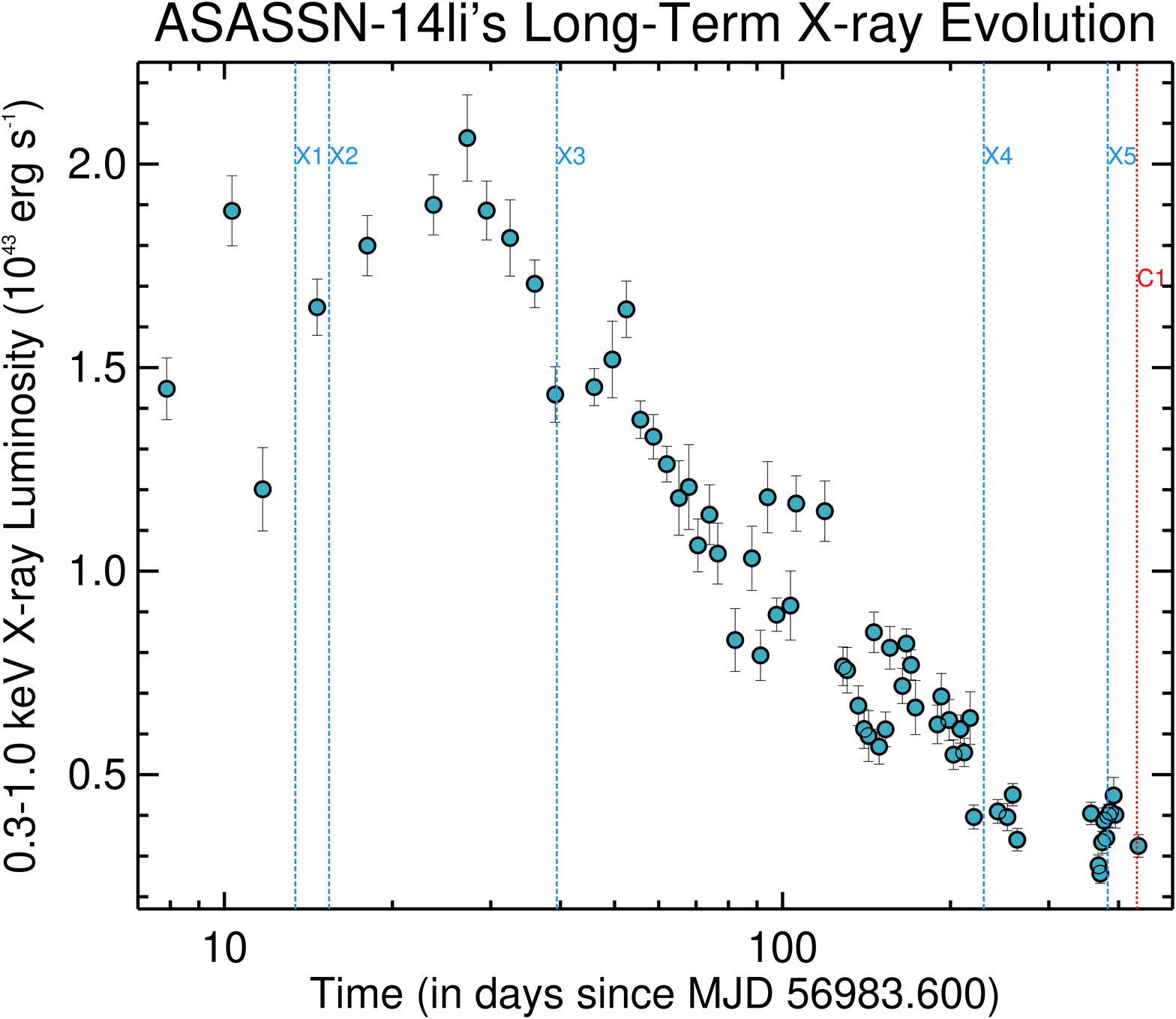}
\end{center}
%\vspace{-.35cm} 
\caption{{\textbf{ASASSN-14li's long-term x-ray light curve.} The data were taken with 
{\it Swift}/XRT (pile-up corrected: see SI). The dashed vertical blue lines marked as X1 to X5 represent 
the five epochs of {\it XMM-Newton} observations. The dotted vertical red line (C1) shows 
the epoch of {\it Chandra}/ACIS observation. %Sudden flux increments, for example, around day 20-30, 40, 100, and 400, suggest that ASASSN-14li could host a massive BH binary (see text and compare with the top-right panel of Fig. 1 of \cite{coughlin17a}). 
}}
\label{fig:swiftlc}
\end{figure}

The BH mass in ASASSN-14li is constrained to lie in between $10^{5.8-7.1} M_\odot$ using well-studied correlations relating BH mass and host galaxy properties\cite{Wevers17,vanvelzen16,Holoien16}. Importantly for our purposes, the observed x-ray energy spectrum is blackbody-like (thermal)\cite{miller15,brownjs17,pasham17} with peak 0.3-1.0 keV luminosity of roughly a few$\times$10$^{43}$ erg/sec (Fig. \ref{fig:swiftlc}). The inferred size of the thermal x-ray emitting region ($\sim$10$^{12}$ cm) is only a few gravitational radii\cite{miller15} and remains roughly constant with time\cite{miller15,brownjs17}. This strongly suggests that x-rays from ASASSN-14li originate from an inner accretion flow close to the BH.

\begin{figure}[ht]
\begin{center}
%\hspace{-1.35cm}
\includegraphics[width=6.5in, angle=0]{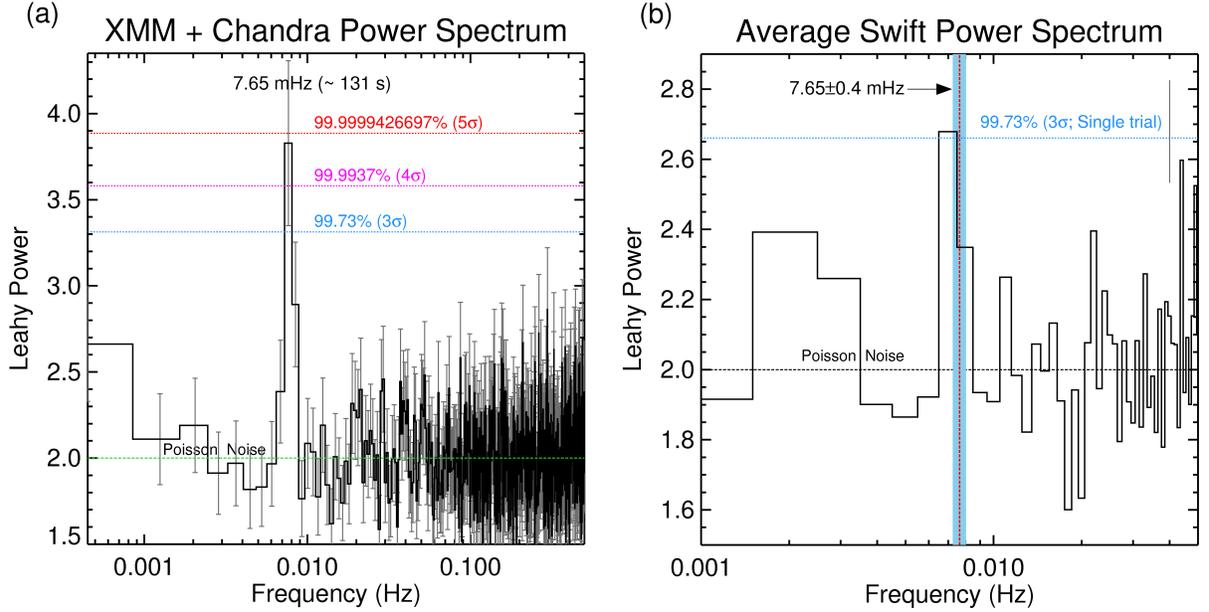}
\end{center}
%\vspace{-.35cm} 
\caption{{\textbf{ASASSN-14li's x-ray QPO at 7.65 mHz is detected by three different telescopes: {\it XMM-Newton}, {\it Chandra} and {\it Swift}.} {\bf (a)} ASASSN-14li's averaged x-ray PDS using eight continuous 10,000 s light curves taken with {\it XMM-Newton} and {\it Chandra}/ACIS. The frequency resolution is 0.8 mHz. The strongest feature in the power spectrum lies at a frequency of 7.65$\pm$0.4 mHz ($\approx$131-seconds). The dashed horizontal blue, magenta, and red lines represent the 3, 4, and 5$\sigma$ white noise statistical contours. The powers surrounding the QPO feature are consistent with white noise (see SI) but we also estimated the QPO significance under various red noise models to find that it is significant at at least the 3.9$\sigma$ level (see SI). $\pm$1$\sigma$ uncertainties are shown in grey. Independently, the QPO is evident at the $\approx$4$\sigma$ and $\gtrsim$ 2.6$\sigma$ levels in the {\it XMM-Newton} and {\it Chandra}/ACIS, respectively (see SI Fig. \ref{fig:separate_pds}). {\bf (b)} Average Swift/XRT PDS. {\it Swift}/XRT's effective area is roughly 1/20$^{\rm th}$ that of {\it XMM-Newton}. Nevertheless, we evaluated the average XRT PDS using the entire archival data. We used 85 continuous 1000 s light curves with a frequency resolution of 1 mHz. The horizontal line shows the 3$\sigma$ contour assuming a single trial search at 7.65 mHz. The highest peak in the power spectrum is at 7.0$\pm$0.5mHz and is consistent with the most prominent feature in the {\it XMM-Newton} and the {\it Chandra} power spectra (SI Fig. \ref{fig:separate_pds}).} }%{\bf (c)} ASASSN-14li's folded X-ray light curve using {\it Chandra} (C1) data. The fold period during that epoch was estimated by oversampling the light curve (see Supplement) to be 134.6$\pm$2.5 s (or 7.43$\pm$0.14 mHz). The best-fitting sinusoidal (dashed red) curve implies a fractional amplitude of 49$\pm$11\% which is consistent with estimate from the PDS (top-left panel). The zero phase is arbitrary. Two cycles are shown for clarity.}
\label{fig:xmmchan_swift_pds}
\end{figure}

In stellar-mass BHs, a sudden onset of accretion often excites quasi-periodic oscillations (QPOs) in the x-ray flux\cite{mcclintockremillard06}. In instances where the majority of the x-ray emission is dominated by the accretion disk, observed QPO frequencies have provided BH spin measurements\cite{franchini17,motta14}. We searched for a stable QPO in the soft x-ray band (0.3-1.0 keV) of ASASSN-14li by combining all its publicly-available data. We extracted an average power density spectrum (PDS) using both {\it XMM-Newton} and {\it Chandra} data taken in six different epochs during the first 450 d after ASASSN-14li's discovery (see Fig. \ref{fig:swiftlc} for its long-term x-ray evolution). The combined x-ray PDS shows a strong feature at 7.65$\pm$0.4 mHz (131-seconds; coherence, $Q$ = centroid-frequency/QPO's-width=16$\pm$6). The QPO is statistically significant at roughly the 4.8$\sigma$ level for a search at all frequencies (trials) below 0.5 Hz (see Fig. \ref{fig:xmmchan_swift_pds}, left panel). Under the assumption that the underlying noise is red, i.e., noise scales inversely with frequency, a conservative lower limit on the statistical significance (false alarm probability) is 3.9$\sigma$ (or 10$^{-4}$; see supplement information, SI).

\begin{figure}[ht]
\begin{center}
%\hspace{-1.35cm}
\includegraphics[width=3.5in, angle=0]{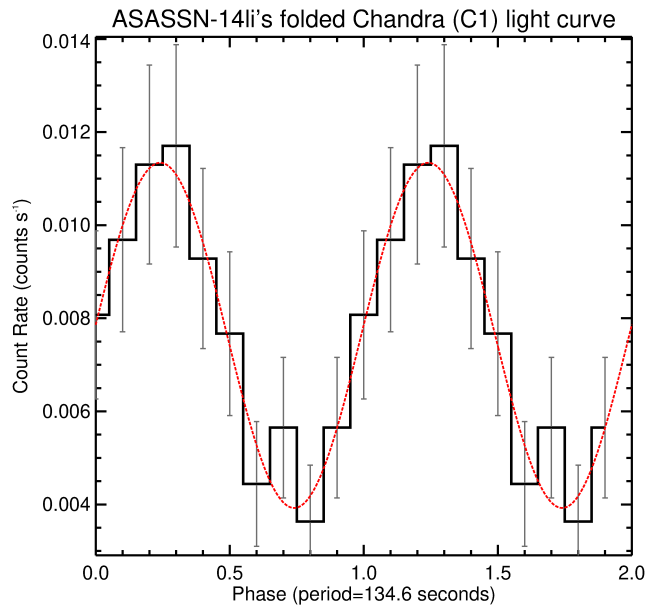}
\end{center}
%\vspace{-.35cm} 
\caption{{\textbf{ASASSN-14li's folded x-ray light curve using {\it Chandra} (C1) data.}} The fold period during epoch C1 was estimated by oversampling the light curve (see SI) to be 134.6$\pm$0.1 s (or 7.43$\pm$0.006 mHz). The best-fitting sinusoidal (dashed red) curve implies a {\bf fractional amplitude of 35$\pm$8\% which is consistent (within the 90\% confidence limits) with the estimate from the PDS (see Fig. \ref{fig:separate_pds})}. The zero phase is arbitrary and two cycles are shown for clarity. $\pm$1-$\sigma$ uncertainties are shown as grey bars. {\bf Figs. \ref{fig:xmmfolds} and \ref{fig:rmsvstime} show the folded {\it XMM-Newton} light curves and the evolution of the QPO's rms amplitude, respectively.} }
\label{fig:foldedchandra}
\end{figure}

 Moreover, the QPO is independently detected in the {\it XMM-Newton} and {\it Chandra} data with a global significance of $\approx$4$\sigma$ and $\gtrsim$2.6$\sigma$, respectively, for a search including all frequencies/trials below 0.5 Hz (see SI, Fig. \ref{fig:separate_pds}). We estimated the QPO's fractional root-mean-squared (rms) amplitude during epoch X5 to be 4$\pm$1\% (see SI). Because pile-up was significant in the first four {\it XMM-Newton} observations similar measurements could not be made. On the other hand, the {\it Chandra} observation was made roughly 420 d after the discovery, by which time ASASSN-14li's flux had declined by roughly a factor of $\approx$ 10, and the pile-up was minimal (see SI). The QPO's fractional rms amplitude in {\it Chandra} data was 59$\pm$11\% (see Fig. \ref{fig:foldedchandra} and SI). This suggests that between X5 and C1, separated roughly by 50 days, the fractional rms amplitude of the QPO amplified by at least an order of magnitude. After establishing the QPO at 7.65 mHz, we also constructed an average {\it Swift} x-ray (0.3-1.0 keV) PDS. The strongest feature in the average {\it Swift} PDS is at 7.0$\pm$0.5 mHz and is consistent with the QPO detected in the {\it XMM-Newton} and {\it Chandra} datasets (Fig. \ref{fig:xmmchan_swift_pds}, right panel).

A {\it Chandra} image shows only a single x-ray point source spatially coincident with the galaxy PGC043234 (SI Fig. \ref{fig:EDF2}). This allows us to assert that the QPO does not originate from a nearby contaminating source. Furthermore, the QPO is detected by three different x-ray detectors. This establishes that the QPO is not an instrumental artifact and is indeed associated with ASASSN-14li. In addition, the supplement movie shows that the QPO signal improves gradually as more PDS are averaged. This implies that the QPO does not originate from a single epoch observation but is present throughout at least the first 450 d of the event. The average {\it Swift} PDS using data acquired over 500 d, the {\it Chandra} PDS from roughly day 420, and the average {\it XMM} PDS all confirm that the QPO is present throughout the first 450 d of the outburst. This implies that the QPO is remarkably stable for 3$\times$10$^{5}$ cycles ($\approx$ 450 d/131 s). While the stability and coherence of the QPO is comparable to the soft-state QPOs of stellar-mass BHs, a modulation amplitude of $>$50\% (Fig. \ref{fig:foldedchandra}) is unprecedented (e.g., \cite{homan01_qpos}).

An alternative scenario in which the oscillation might be a neutron star pulsation is extremely unlikely for many reasons: large x-ray, optical/UV and radio photospheric sizes\cite{pasham17,Holoien16,brownjs17,pashamsjort17}, high persistent bolometric luminosity\cite{pasham17,brownjs17}, and a very soft x-ray spectrum\cite{miller15,brownjs17}. In general, the multiwavelength properties of ASASSN-14li are strikingly similar to many previously-known TDEs, and are unlike any known neutron star outburst (see SI for more details). 

It is known empirically that the masses of central BHs are correlated with the properties of their host galaxies\cite{gultekin09,mcconnellma13}. For instance, the velocity dispersion of stars in the inner bulges of galaxies ($\sigma$) is correlated with the BH mass ($M$), and is commonly referred to as the $M$-$\sigma$ relation. In addition, the total stellar mass in the bulge and the optical luminosity of the host galaxy are shown to correlate with the BH mass\cite{mcconnellma13}. These empirical relations suggest that the BH in ASASSN-14li has a mass anywhere in between 10$^{5.8-7.1}$ $M_{\odot}$\cite{vanvelzen16,Wevers17,Holoien16}. This value is also consistent with mass derived independently from detailed physical modeling of ASASSN-14li's multi-waveband (x-ray, optical and UV) light curves \cite{miller15}. Assuming this BH mass range, we compared the 7.65 mHz QPO frequency to the five possible frequencies of motion of a test particle orbiting a spinning BH in Kerr spacetime\cite{bardeen72,motta14}. 

\begin{figure}[ht]
\begin{center}
%\hspace{-1.35cm}
\includegraphics[width=4.5in, angle=0]{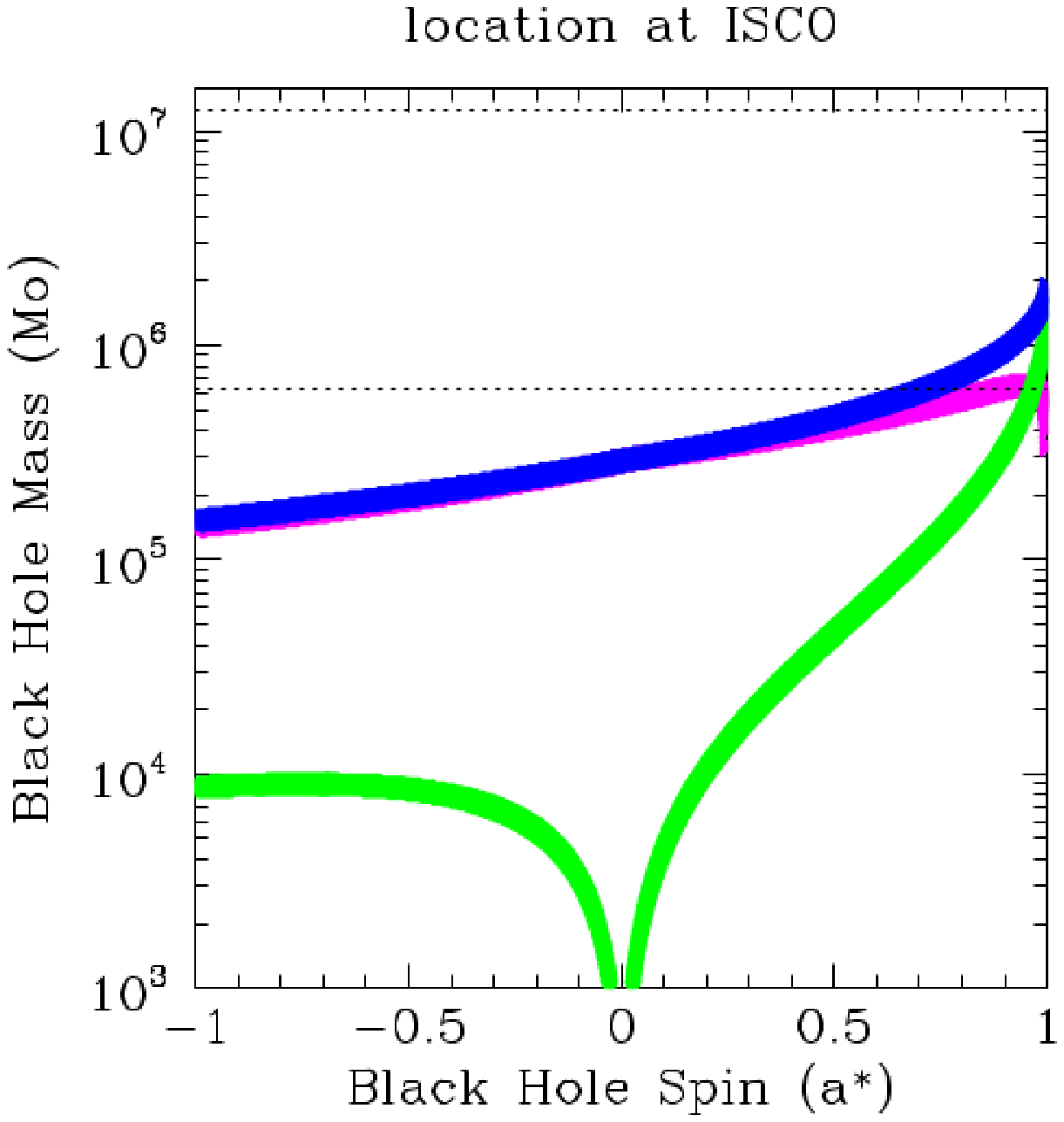}
\end{center}
%\vspace{-.35cm} 
\caption{{\textbf{Black Hole dimensionless spin parameter vs mass contours:} Spin vs mass contours assuming the 7.65 mHz QPO is associated with any of three particle frequencies: Keplerian frequency ($\nu_{\phi}$, blue), vertical epicyclic frequency ($\nu_{\theta}$, magneta) and Lense-Thirring precession ($\nu_{\phi}$ - $\nu_{\theta}$, green) at the innermost stable circular orbit (ISCO). At the ISCO the radial epicyclic frequency ($\nu_{r}$) is zero and the periastron precession frequency ($\nu_{\phi}$ - $\nu_{r}$) is thus equal to the Keplerian frequency (see SI). The widths of these contours reflect the  QPO's width of 0.7 mHz (upper limit). The dotted horizontal lines show ASASSN-14li's BH mass range (10$^{5.8-7.1}$ $M_{\odot}$) estimated from its host galaxy scaling relations. Within this mass range, the only formal solutions are the ones that require the BH spin to be greater than 0.7.}}
\label{fig:massspincontours}
\end{figure}

All five particle frequencies are determined by the BH's mass, spin and the radial distance of the emitting region. In soft-state stellar-mass BHs the inner edges of accretion disks extend to a constant radius for a wide range in accretion rates (e.g., \cite{jacklmc10}). The natural inner radius predicted from Einstein's theory of general relativity is the so-called innermost stable circular orbit (ISCO), which depends to leading order on BH spin. Because ASASSN-14li also appears to be disk-dominated we started our frequency comparison using the ISCO as the radial distance (Fig. \ref{fig:massspincontours}). Surprisingly, even with the closest possible location, i.e, the ISCO, the only solutions are the ones that require the BH to be rapidly spinning. A lower limit on BH's dimensionless spin parameter ($a^{*}$)\footnote{$a^{*}$ = $Jc/GM^{2}$, where $J$ is BH's angular momentum.} can be calculated from the BH spin vs mass contours (Fig. \ref{fig:massspincontours}). This corresponds to the intersection of the BH mass lower limit and the fastest frequency, which at any given radius is the Keplerian frequency (blue in Fig. \ref{fig:massspincontours}). This implies that ASASSN-14li's spin is greater than 0.7 (Fig. \ref{fig:massspincontours}). Choosing any larger radius will only push this limit to a higher spin value. 

Many researchers expect that a proper theory for disk oscillations must consider normal mode analyses for a detailed accretion flow in strong gravity. If we ignore modes with frequencies higher than the azimuthal (Keplerian) frequency (but see below), then we can interpret Fig. \ref{fig:massspincontours} as describing a lower limit to the spin (e.g., \cite{stroh01}) of the MBH that caused the TDE. Alternatively, we can interpret the figure as an upper limit of $2\times10^6 M_\odot$ on the black hole mass for a maximum astrophysically plausible spin of a*=0.998\cite{thornemaxspin}.

On the other hand, it is possible that ASASSN-14li's host galaxy and the disrupting BH may not obey the empirical scaling laws\cite{Greene_msimga_fail_lowmass} and instead the BH mass could be significantly below a value of a few$\times$10$^{5}$ $M_{\odot}$. If that were the case, then the BH could also have a moderate spin. But, more interestingly, it would imply that the disrupting BH is a member of the elusive class of intermediate-mass black holes whose existence has so far been controversial.

In any case, such a high-amplitude quasi-periodic phenomenon that is stable for years has never been seen before from any BH. This QPO has the highest dimensionless (i.e., in units of $c^3/GM$) frequency ever seen from a BH accretion disk. This implies that the radiation producing this QPO originates from very near the BH's event horizon, and clearly rules out alternative models for x-ray radiation that require an emitting region far away from the black hole. The extreme phenomenon of TDEs is relatively new and we are only beginning to understand their underlying physics as more sensitive observations are becoming available. This discovery presents a new probe into the complex accretion physics of a star getting disrupted by a MBH and may require a new physical mechanism to explain its origin (see SI for some discussion).

In addition to its remarkable stability, high amplitude and fast timescale, there are other peculiarities when compared to x-ray QPOs in accreting stellar-mass black holes. For example, the high-frequency QPOs (frequencies of a few$\times$100 Hz) of accreting stellar-mass black holes are seen only in hard x-rays ($>$2keV)\cite{mcclintockremillard06} and not in soft states\cite{homan01_qpos}. By contrast, ASASSN-14li's energy spectrum is very soft (9). {\bf The rapid rise in QPO's rms amplitude is also unprecedented}. Given this range of surprising properties ASASSN-14li's QPO may represent a disk oscillation mode never seen in other systems, and thus it may not be strictly valid to compare it directly with known QPOs of stellar-mass black holes.

Previously, Reis et al. (2012)\cite{Reis12} reported the first case of a quasi-periodicity (at $\approx$200-seconds) from a TDE SwJ1644+57. However, SwJ1644+57 is an atypical TDE where the entire electromagnetic radiation was dominated by a jet directly pointing along our line of sight (e.g., \cite{Bloom11}). Radio followup indicates that only a small fraction of thermal TDEs launch collimated jets \cite{generozov17}, and obviously only a small fraction of such jets would align with our line of sight. Moreover, compared to ASASSN-14li, SwJ1644+57's periodicity was roughly 15 times weaker in amplitude and was present only for a short duration of at most a few weeks after its discovery. Previously, it had been unclear if standard TDEs could exhibit any quasi-periodic variability. Our discovery of a very loud (fractional {\bf rms$\sim$50\%}), and more importantly very persistent (stable over years) quasi-periodicity from a poster-child TDE suggests that QPOs may be common in TDEs.
%all TDEs may, in principle, exhibit such phenomenon. 

In summary, fast x-ray variations (QPOs) originate from the strongest gravity regime in the immediate vicinity of BHs. Combining this with the fact that the centroid period of the QPO is stable then strongly suggests that the QPO is tied to the fundamental properties, viz., the mass and the spin of the BH at the heart of the disruption. Similar QPOs in future TDEs will not only probe general relativity in the strong regime but may also allow us to build a census of MBHs in the Universe.

%%%%%%%%%%%%%%%%%%%%%%%%%%%%%%%%%%%%%%%%%%%%%%%%%%%%%%%%%%%%%%%%%%%%%%%%%%%%%%%%%%%%%%%%%%%%%%
% ---- Figure--- Figure ---- Figure--- Figure ---- Figure--- Figure --- Figure--- Figure ----%
%%%%%%%%%%%%%%%%%%%%%%%%%%%%%%%%%%%%%%%%%%%%%%%%%%%%%%%%%%%%%%%%%%%%%%%%%%%%%%%%%%%%%%%%%%%%%%

\newpage
\vfill\eject

%%%%%%%%%%%%%%%%%%%%%%%%%%%%%%%%%%%%%%%%%%%%%%%%%%%%%%%%%%%%%%%%%%%%%%%%%%%%%%%%%%%%%%%%%%%%%%
% ---- Figure--- Figure ---- Figure--- Figure ---- Figure--- Figure --- Figure--- Figure ----%
%%%%%%%%%%%%%%%%%%%%%%%%%%%%%%%%%%%%%%%%%%%%%%%%%%%%%%%%%%%%%%%%%%%%%%%%%%%%%%%%%%%%%%%%%%%%%%

\newpage
\vfill\eject

%%%%%%%%%%%%%%%%%%%%%%%%%%%%%%%%%%%%%%%%%%%%%%%%%%%%%%%%%%%%%%%%%%%%%%%%%%%%%%%%%%%%%%%%%%%%%%
% ---- Figure--- Figure ---- Figure--- Figure ---- Figure--- Figure --- Figure--- Figure ----%
%%%%%%%%%%%%%%%%%%%%%%%%%%%%%%%%%%%%%%%%%%%%%%%%%%%%%%%%%%%%%%%%%%%%%%%%%%%%%%%%%%%%%%%%%%%%%%

\newpage
\vfill\eject

%%%%%%%%%%%%%%%%%%%%%%%%%%%%%%%%%%%%%%%%%%%%%%%%%%%%%%%%%%%%%%%%%%%%%%%%%%%%%%%%%%%%%%%%%%%%%%
% ---- Figure--- Figure ---- Figure--- Figure ---- Figure--- Figure --- Figure--- Figure ----%
%%%%%%%%%%%%%%%%%%%%%%%%%%%%%%%%%%%%%%%%%%%%%%%%%%%%%%%%%%%%%%%%%%%%%%%%%%%%%%%%%%%%%%%%%%%%%%
\newpage
\vfill\eject

\singlespace
\begin{scilastnote}
\item[] {\bf Acknowledgements.} Pasham would like to thank David Huenemoerder, Adam Ingram, Phil Uttley, and Mike Nowak for valuable discussions. This work is based on observations made with {\it XMM-Newton}, {\it Chandra}, and {\it Swift}. {\it Swift} is a mission that is managed and controlled by NASA's Goddard Space Flight Center (GSFC) in Greenbelt, Maryland, USA. The data used in the present article is publicly available through NASA's {\it HEASARC} archive. 
\item[]  All the data presented here is public and can be found in the ESA/{\it XMM-Newton}, NASA/{\it Chandra} and {\it Swift} archives: \\
http://nxsa.esac.esa.int/nxsa-web/\\
http://cxc.harvard.edu/cda/\\
http://heasarc.nasa.gov/docs/swift/archive/\\
%\item[] {\bf Supplement Material} \\
%Figs. S1 to S9\\ 
%References (31-67) \\
\end{scilastnote}

%%%%%%%%%%%%%%%%%%%%%%%%%%%%%%%%%%%%%%%%%%%%%%%%%%%%%%%%%%%%%%%
%
%SUPPLMENT
%
%%%%%%%%%%%%%%%%%%%%%%%%%%%%%%%%%%%%%%%%%%%%%%%%%%%%%%%%%%%%%%%%
\newpage
\newpage

\setcounter{page}{1}
\renewcommand{\theequation}{S\arabic{equation}}
\renewcommand{\thefigure}{S\arabic{figure}}
\renewcommand{\thetable}{S\arabic{table}}
\setcounter{figure}{0}

\singlespace

%\noindent{\Huge {\bf Supplement}} \\[10 pt]

\noindent{\Huge {\bf Supplementary Information}} \\[10 pt]
\section{\Large{\bf Data Reduction. }}\label{supsec:data}
The data used in this work has been acquired by three different x-ray 
telescopes: {\it Swift}, {\it XMM-Newton}, and {\it Chandra}. The X-Ray 
Telescope (XRT) on board {\it Swift} started monitoring ASASSN-14li 
roughly a week after its discovery by the All-Sky Automated Survey for 
Supernovae (ASAS-SN\cite{shappee14}) on MJD 56983.6\cite{Holoien16}. 
Since discovery on 22 November 2014 until May 2017, {\it Swift} observed 
ASASSN-14li on over 100 occasions with each observation lasting between 
a few$\times$(100-1000) seconds. {\it XMM-Newton} and {\it Chandra}--with 
effective areas larger than the XRT--provided fewer but more sensitive 
observations each lasting anywhere between 10,000 and 90,000 seconds.

We started our analysis with {\it Swift}/XRT data to assess the long-term 
x-ray evolution as follows. As noted by earlier works\cite{miller15,pasham17}, 
the individual XRT data sets are indeed piled-up. To mitigate the effect of 
pile-up we extracted event lists from an annulus region centered on the 
source by excluding an inner pile-up radius in each individual observation 
by following the procedure outlined by the XRT pile-up guide\footnote{http://www.swift.ac.uk/analysis/xrt/pileup.php}. The 
individual XRT observations only have a few counts and thus cannot be used 
to constrain the spectral shape. Therefore, we extracted average energy 
spectra by combining neighboring observations until a total of $\sim$3500 counts 
were reached. Similar to earlier works\cite{miller15} we modeled each 
energy spectrum with an absorbed black body modified by red-shift of the 
host galaxy and implemented it in the x-ray spectral fitting package, 
XSPEC\cite{arnaud96}, as {\tt phabs*(zashift(phabs*bbodyrad)}. We 
estimated the flux and thus the luminosity in each individual {\it Swift} 
observation by fitting it with the same black body function but limited the 
column density and the disk temperature values to the nearest (in time) 
averaged spectral values (see \cite{pashamsjort17} for more specific details and the best-fit model parameters). 
ASASSN-14li's final XRT x-ray (0.3-1.0 keV) long-term light curve is shown 
in Fig. \ref{fig:swiftlc}.

{\it XMM-Newton} and {\it Chandra} observed ASASSN-14li on multiple 
occasions with six and three data sets publicly available at the time of 
writing of this paper. However, one of the {\it XMM-Newton} observations 
(obsID: 0770980501) was severely effected by background flaring and two of 
the {\it Chandra} data sets carried out with the High-Resolution Camera 
(HRC) were background dominated. Therefore, we did not consider these data 
sets for further analysis. After this initial screening we were left with 
five {\it XMM-Newton} and one {\it Chandra} observation. The vertical 
lines in Fig. \ref{fig:swiftlc} mark the epochs of these observations. A 
summary of these observations can be found in Table. \ref{table:table1}.

\begin{figure}[ht]
 \begin{center}
 %\vspace{-0.75cm}
 \includegraphics[width=4.5in, angle=0]{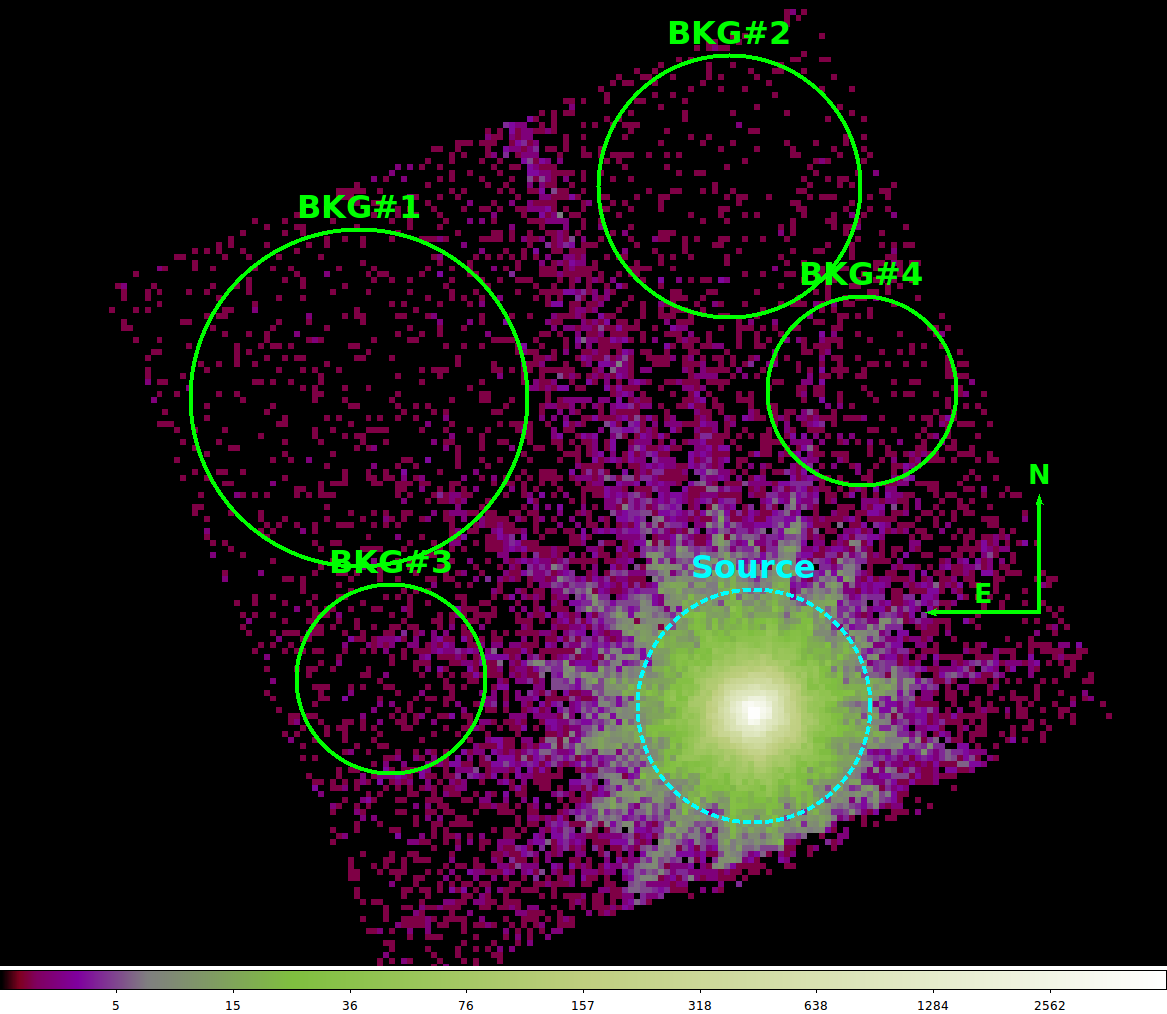}
 \end{center}
 %\vspace{-.35cm} 
 \caption{{\bf A sample {\it XMM-Newton} EPIC (pn+MOS) 0.3-1.0 keV image of ASASSN-14li's field of view.} The source extraction region is indicated by a dashed cyan circle while the background extraction regions are shown as green circles. The north and east arrows are each 40'' long. For this particular image we used the data set with an obsID of 0722480201.}
 \label{fig:xmmepicimage}
 \end{figure}

We used the {\it XMM-Newton} Standard Analysis System ({\tt xmmsas}: 
version 15.0.0) to extract the images and the event lists from all the 
five {\it XMM-Newton} data sets. Because the detection sensitivity of a 
quasi-periodic feature in the light curves increases sharply with the 
count rate\cite{vanderklis88}, we combined the data acquired by all the 
three detectors (pn, MOS1 and MOS2) on board the European photon imaging 
camera (EPIC) and considered only those epochs during which all the 
three detectors were operating. We started our analysis by 
reducing/reprocessing the datasets to extract the level-2 event-lists. 
We first extracted images of ASASSN-14li's field of view and visually 
confirmed that there are no contaminating sources nearby. We extracted 
source events from a circular region of radius 40'' centered on the 
source. All the observations were taken in a small window mode to enable 
faster readout. To better constrain the background variability, events 
were extracted from four circular regions offset from the source 
and with radii of 58'', 45'', 32.5'', and 32.5''. A sample {\it XMM-Newton}/EPIC 
(pn+MOS) image is shown in Fig. \ref{fig:xmmepicimage} highlighting the source and 
the background extraction regions. A significant fraction of these data 
sets were affected by background flaring. Because these high-amplitude 
background flux variations can sometimes manifest as quasi-periodic 
features in the power spectra we carefully removed these high background 
flux epochs from our analysis. This--combined with our requirement to 
consider only times when all the three (pn+MOS1+MOS2) detectors were active--resulted in a number of Good Time 
Intervals (GTIs) in each individual observation (see left panels of Figs. 
\ref{fig:xmmlcs1} and \ref{fig:xmmlcs2} below).

\begin{figure}[ht]
 \begin{center}
 \vspace{-1cm}
 \includegraphics[width=6in, angle=0]{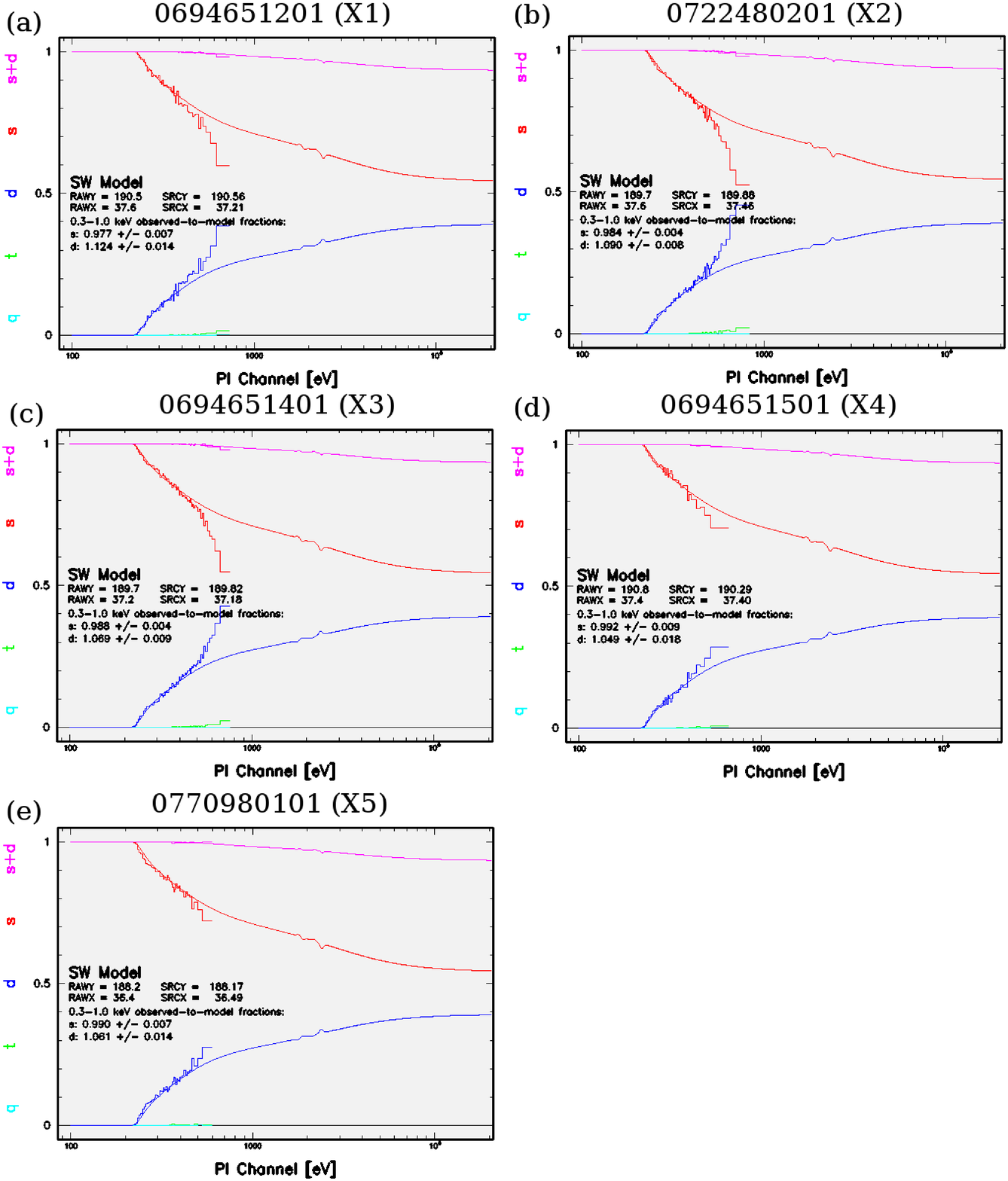}
 \end{center}
 \vspace{-.5cm} 
 \caption{{\bf ASASSN-14li's {\tt epatplot} outputs showing photon pile-up in {\it XMM-Newton} observations.} 
In each panel the histogram show the observed distribution of the single (s; blue), double (d; magenta),  triple (t; green) and quadruple (q; cyan) pixel events. The solid curves are the expected distributions. It is obvious that the expected distributions, especially for the single and double events, do not agree well with the observed data. This mismatch is an indication of photon pile-up which is evident in all the observations with X5 being least effected {\bf (see sec. \ref{supsec:data} for more discussion)}.   }
 \label{fig:xmmpileupplots}
 \end{figure}

Earlier studies\cite{miller15, kara2018} have found that ASASSN-14li is 
piled-up even in the {\it XMM-Newton} observations. Following the standard 
procedure to check for pileup, as outlined in the {\it XMM-Newton} data analysis 
guide\footnote{https://www.cosmos.esa.int/web/xmm-newton/documentation}, we 
also reach the same conclusion. Fig. \ref{fig:xmmpileupplots} shows the output from the 
{\tt xmmsas} task {\tt epatplot} for the EPIC-pn detector for all the five 
observations (Xn, n from 1 to 5 as marked in Fig. \ref{fig:swiftlc}). {\bf The plots show the migration of X-ray events to higher patterns, i.e., from single to double, because of pile-up. The underlying reason is that when pile-up occurs the detector incorrectly interprets multiple single pixel events in adjacent pixels as a single multi-pixel event. This results in a deficit of single pixel events and an excess of double (or higher) pixel events as seen in Fig. \ref{fig:xmmpileupplots}. Furthermore, for a given detector, the expected fraction of total X-ray events that create a charge cloud pattern within $i$ (= single, double, etc) pixels usually depends on the energy as can be seen from the solid curves in Fig. \ref{fig:xmmpileupplots}.} The obvious disagreement 
between the expected (solid curves) and the observed (histograms) distributions 
of the single and the double pixel events suggests that the observations are 
indeed piled-up. It is interesting to note that while the observed count rates 
are well below the pileup threshold count rates given in the {\it XMM-Newton} 
calibration documents\footnote{http://xmm2.esac.esa.int/docs/documents/CAL-TN-0200-1-0.pdf}, 
pile up does occur (Fig. \ref{fig:xmmpileupplots}), and this may be due to the extreme soft 
nature of ASASSN-14li\cite{kara2018}.

For the purposes of variability studies, pile-up can have two major effects: 
(1) it reduces the overall count rate, and (2) it may reduce the fractional 
root mean-squared (rms) amplitude\footnote{http://cxc.harvard.edu/ciao/dictionary/pileup.html}. 
For example, Tomsick et al. (2004)\cite{tomsickpileup} showed via simulations 
that the fractional rms amplitude of piled-up {\it Chandra} x-ray data of the 
transient XTE J1650-500 reduced by roughly 1\%. With regards to detecting a 
periodic/quasi-periodic signal, even though the mean rate is reduced, the 
count rate at the peak of the waveform is reduced slightly more than at the 
{\bf trough}, resulting in a reduced fractional 
rms\footnote{http://cxc.harvard.edu/ciao/download/doc/pileup\_abc.pdf}. To 
alleviate the pile-up issue and at the same time not compromise too much on 
the count rate, we considered all events within a 40'' circular region for 
power spectral analysis.

Because ASASSN-14li's energy spectrum is very soft the count rates in 1-10 keV band are negligible compared to the 0.3-1.0 keV range. We estimated the ratio of 1.0-10.0 keV to 0.3-1.0 keV background subtracted source count rates in the five {\it XMM-Newton} obsIDs 0694651201, 0722480201, 0694651401, 0694651501, and 0770980101 to be 0.009, 0.002, 0.001, 0.003, and 0.005, respectively. In other words, ASASSN-14li's 1-10 keV flux is less than 1\% of 0.3-1.0 keV flux, and is dominated by the background. A similar conclusion was reached by earlier works\cite{brownjs17}. Therefore, we only considered the soft 0.3-1.0 keV events for power analysis. {\it XMM-Newton} data analysis guide\footnote{https://heasarc.gsfc.nasa.gov/docs/xmm/sas/USG/epicpileup.html} recommends the use of single pixel events {\tt PATTERN==0} for piled-up data. Therefore we added an additional filter to only include single pixel events.

\begin{figure}[ht]
 \begin{center}
 \vspace{-0.75cm}
 \includegraphics[width=4.5in, angle=0]{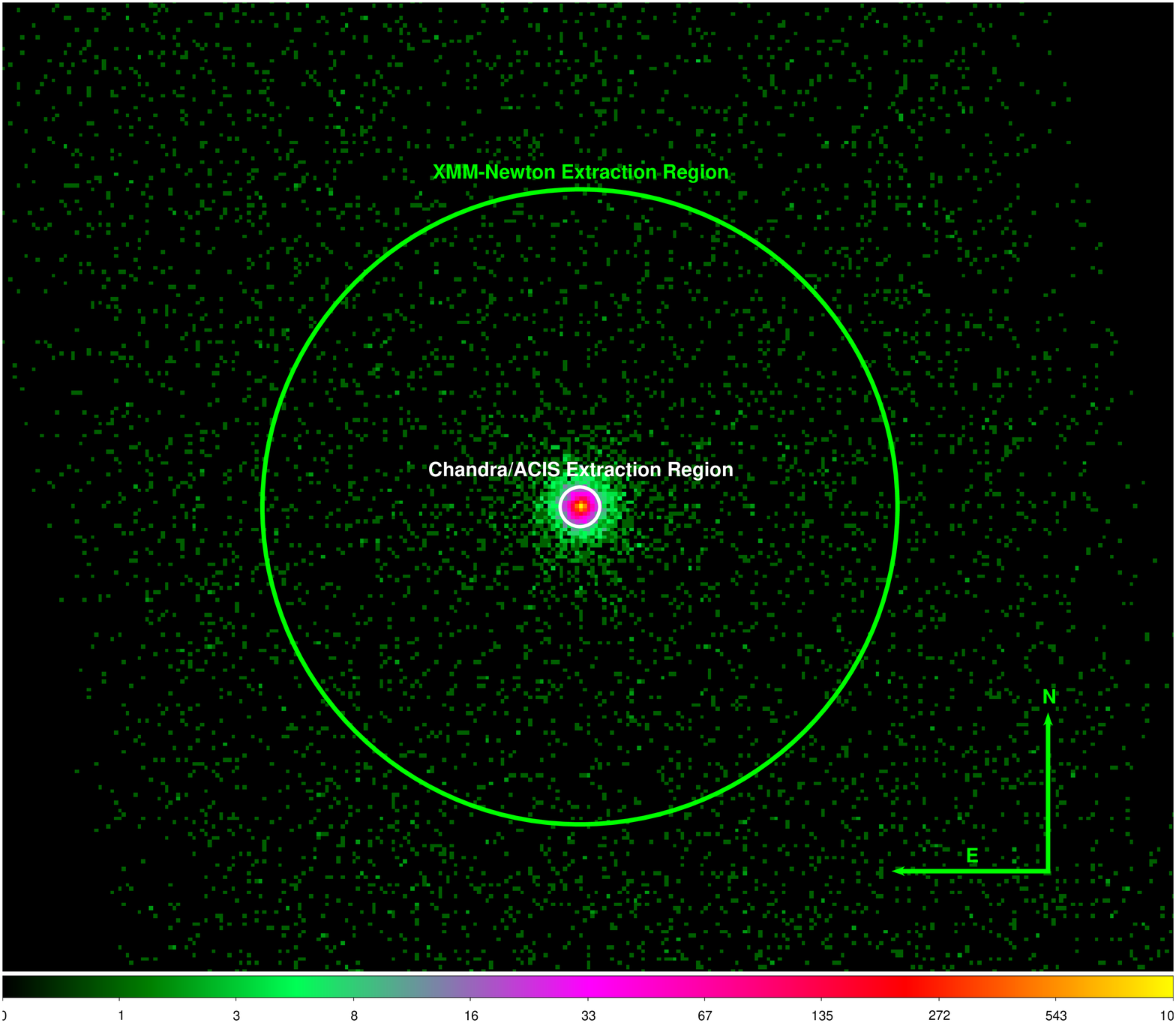}
 \end{center}
 %\vspace{-.35cm} 
 \caption{{\bf Chandra/ACIS x-ray image of ASASSN-14li shows no contaminating sources.} 
It is clear that there is only one source, ASASSN-14li, and no obvious evidence for 
source contamination. {\it Chandra} and {\it XMM-Newton} extraction regions are shown 
as white (2.5'') and green (40'') circles, respectively.}
 \label{fig:EDF2}
 \end{figure}

We then extracted an image and an event list from {\it Chandra}'s 
Advanced CCD Imaging Spectrometer (ACIS) using the Chandra Interactive 
Analysis of Observations tool called {\tt ciao}. Currently, {\it 
Chandra} offers the best spatial resolution in x-rays. We inspected
ASASSN-14li's ACIS image--with a spatial resolution of 0.5''--and found 
that it is the only source in that field of view (see Fig. 
\ref{fig:EDF2}). Source events were extracted from a circular region of radius 2.5'' 
centered on the source while the background events were extracted from 
a region significantly ($\gtrsim$ 3000 times) larger than the source 
region. Again we utilized only single pixel events in the calibrated 
energy range of 0.4-1.0 keV.

\section{\Large{\bf Power Spectral Analysis. }}\label{supsec:psdanal}
\begin{figure}[ht]
 \begin{center}
 \vspace{-1cm}
 \includegraphics[width=6.25in, angle=0]{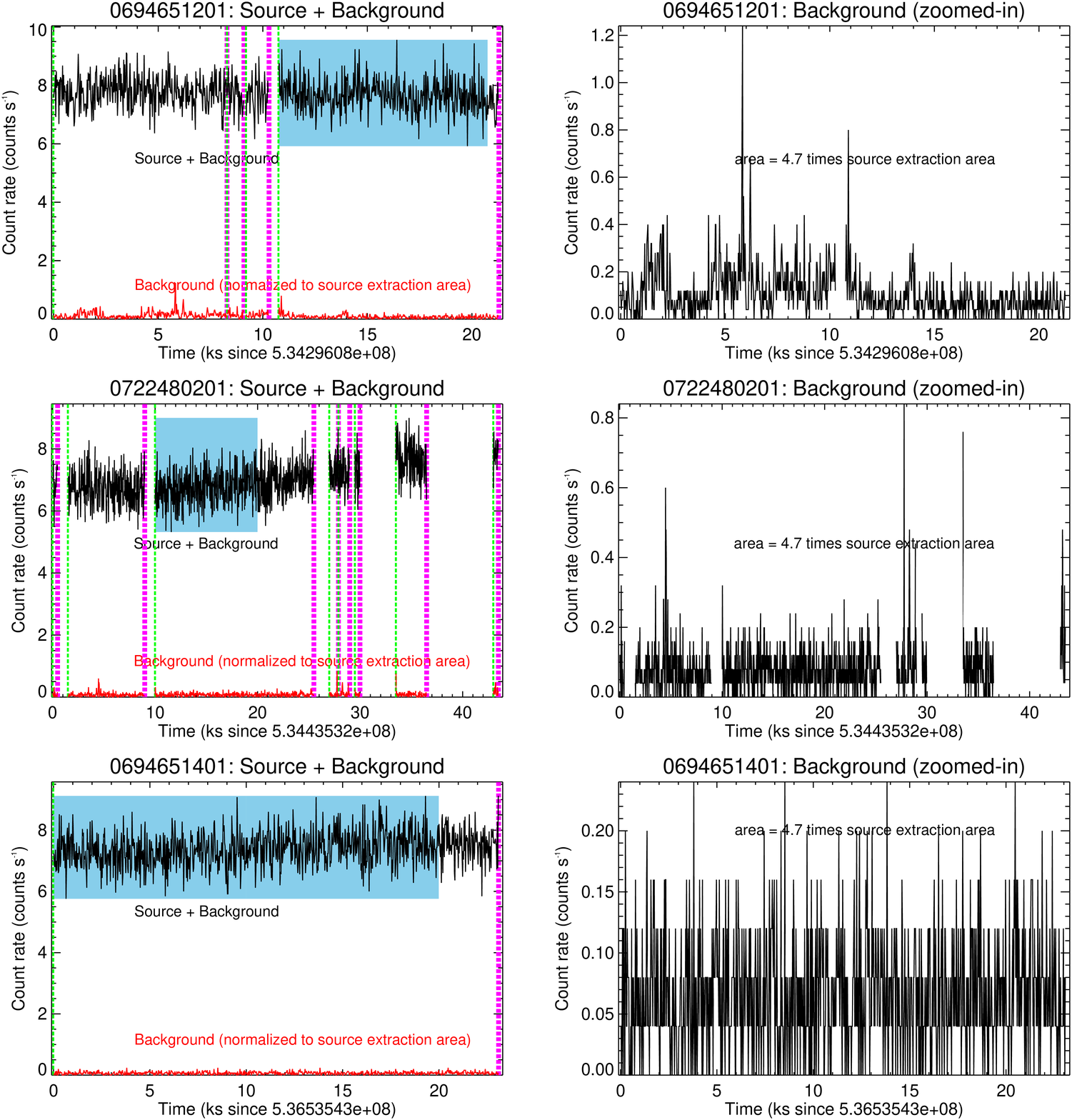}
 \end{center}
 \vspace{-.5cm} 
 \caption{{\bf ASASSN-14li's source and background light curves using {\it XMM-Newton} data sets.} 
{\bf Left panels:} Source + background (black) and background (red) X-ray (0.3-1.0 keV) light curves binned at a resolution of 25 s. The background light curves (red) are normalized to the source extraction area, i.e., divided by a factor of the ratio of the background to source area (=4.7). The green and the magenta mark the beginning and the end of all GTIs greater than 10 s. The first 10 ks (or integer multiple of 10 ks segments) of all GTIs greater than 10 ks are highlighted by a shaded blue rectangle. These are the data segments used for constructing the average PDS in Fig. \ref{fig:xmmchan_swift_pds}. The exact values of the GTIs used for constructing the average PDS are listed in Table. \ref{table:table1}. {\bf Right panels:} Zoom-in on the background light curves. They were extracted from a total area of 4.7 times the source extraction area. }
 \label{fig:xmmlcs1}
 \end{figure}

\begin{figure}[ht]
 \begin{center}
%  \vspace{-0.75cm}
 \includegraphics[width=6.25in, angle=0]{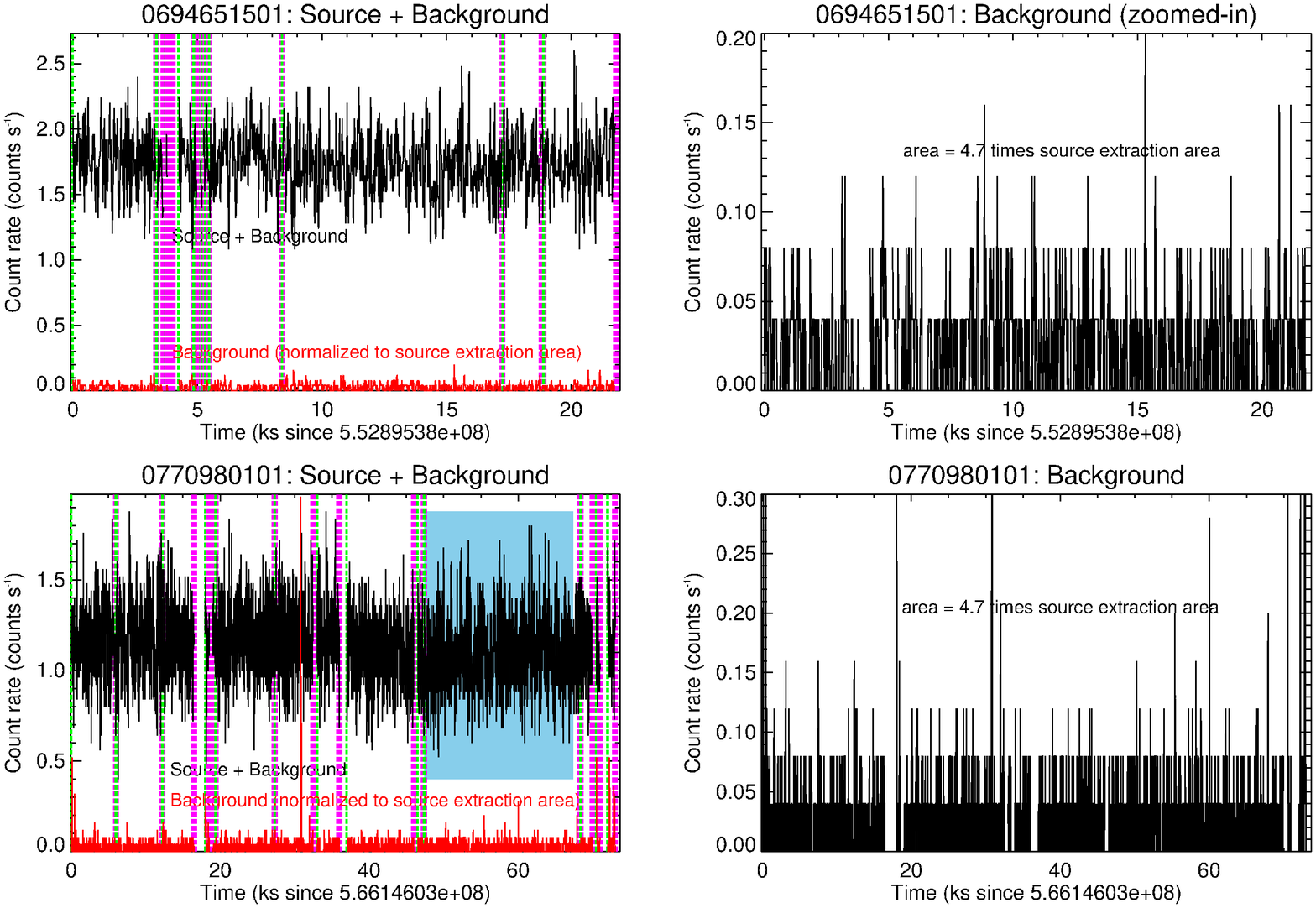}
 \end{center}
 %\vspace{-.35cm} 
 \caption{{\bf Same as Fig. \ref{fig:xmmlcs1} showing the data from other observations. } 
 }
 \label{fig:xmmlcs2}
 \end{figure}

We first divided the data into 10,000-second continuous segments and extracted their light curves with a 1-second time resolution. Because the individual observations (see Table. \ref{table:table1}) were broken into several GTIs this selection resulted in a total of eight uninterrupted data segments. The epochs of these eight 10,000 s light curves, i.e., the start and the end times in units of seconds since modified Julian date (MJD) of 50814.0, are highlighted (blue shade) in Figs. \ref{fig:xmmlcs1} and \ref{fig:xmmlcs2} and are also tabulated in Table \ref{table:table2}. We then constructed a Leahy normalized\cite{leahy83} power density spectrum (PDS)--where the mean Poisson noise level is 2--from each of these 10,000 second light curves. This resulted in eight PDS which were all combined to obtain an average PDS of ASASSN-14li. The most striking feature in the resulting PDS is a quasi-periodic oscillation (QPO) candidate at roughly 7.65 mHz.

\subsection{Underlying Noise Distribution:}
The statistical significance of a power fluctuation in a PDS depends on the underlying distribution of the noise powers. Except at 7.65 mHz ($\pm$ 1 frequency bin), which is dominated by the QPO, the rest of the PDS between 0.001 and 0.5 Hz appears to be roughly constant with no obvious dependence on frequency, i.e., the noise appears to be white. A statistical test for white noise is a test of whether the powers are $\chi^2$ distributed\cite{vanderklis88}. 

\subsubsection{Probability plot:}
\begin{figure}[ht]
 \begin{center}
 %\vspace{-0.75cm}
 \includegraphics[width=6.25in, angle=0]{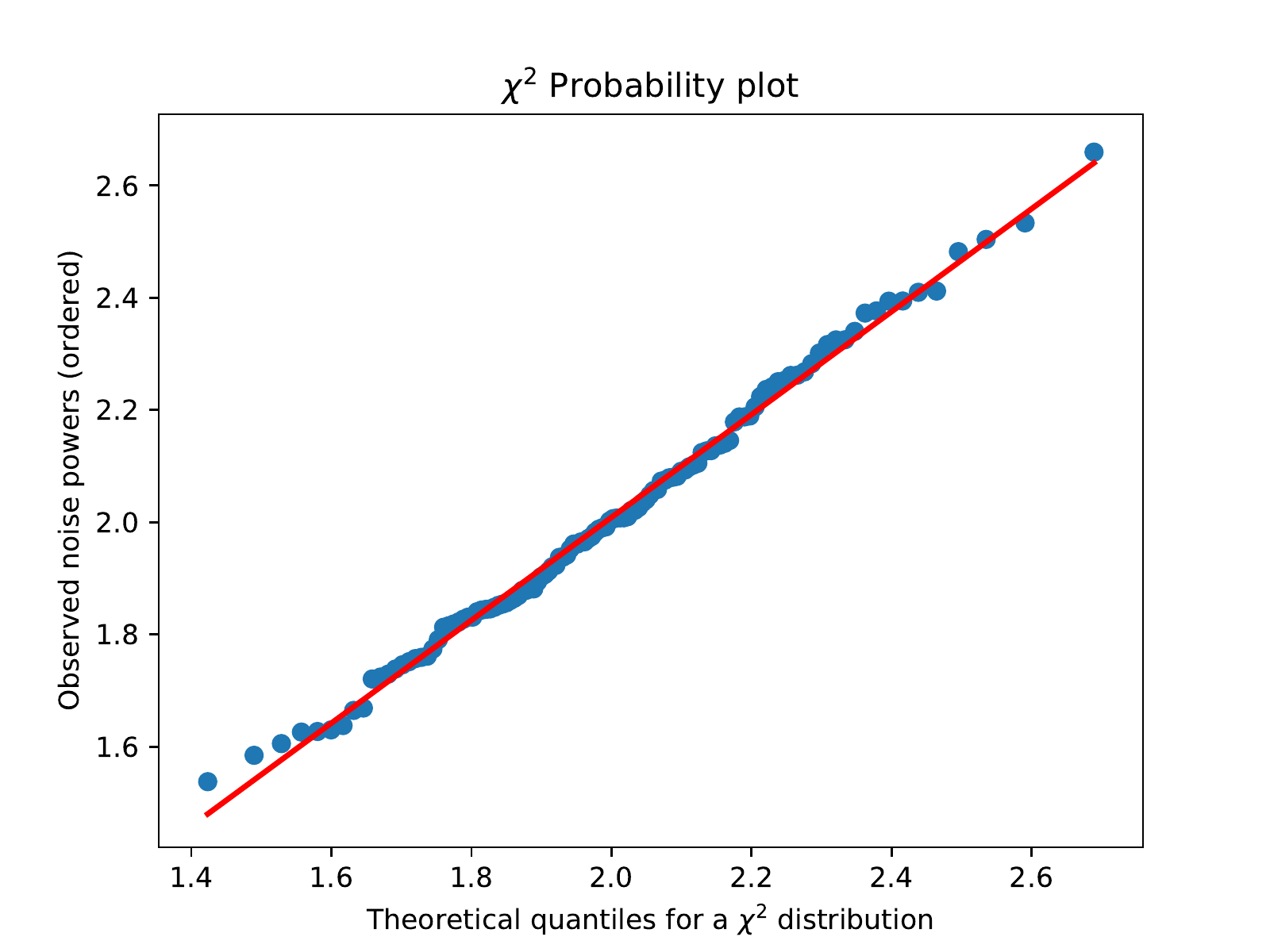}
 \end{center}
 %\vspace{-.35cm} 
 \caption{{\bf Probability plot to visually assess ASASSN-14li's power spectral noise powers against white noise ($\chi^2$ distribution).} If the data points lie on straight line, as they do, then it suggests (at least qualitatively) that the data are consistent with the hypothesized model which in our case is a $\chi^2$ distribution with 128 degrees of freedom scaled by a factor of 1/64 (see text for details).}
 \label{fig:probplot}
 \end{figure}

As a first step, in order to visually assess whether the noise powers in the vicinity of the QPO candidate follow a $\chi^2$ distribution, we constructed a $\chi^2$ probability plot using the noise powers in the frequency bins between 0.001 and 0.1 Hz. We excluded the three bins containing the signal itself at 7.65 mHz. The resulting plot is shown in Fig. \ref{fig:probplot}. A probability plot is a commonly used statistical tool that shows the theoretical quantiles of the assumed distribution against the ordered sample values, i.e., noise powers in our case. We used Filliben's\cite{filliben1975} formula for estimating the theoretical quantiles. In our case the distribution is $\chi^2$ with 2$\times$8$\times$8 degrees of freedom (dof) scaled by a factor of 1/64. This particular $\chi^2$ distribution was used because we averaged in frequency by a factor of 8 and averaged 8 individual power spectra. If the data lie on a straight line then it indicates that they are consistent with the theoretical distribution that it is compared against. It is evident from Fig. \ref{fig:probplot} that the data points on the probability plot follow a straight line (red line) quite well and thus appear consistent with a $\chi^2$ distribution with 128 dof scaled by a factor of 1/64.

\subsubsection{Comparing power-law + constant vs constant models for the continuum.}\label{supsec:modelpsd}
Also, as a straightforward check, we fit the Leahy-normalized power spectrum in the left panel of Fig. \ref{fig:xmmchan_swift_pds} with a constant plus a Lorentzian for the QPO and compared the improvement in $\chi^2$ by adding a powerlaw. The former yielded a $\chi^2$ of 98 for 121 dof while a model with powerlaw resulted in a $\chi^2$ of 94 for 119 dof. Repeating the same exercise on the unbinned PDS, i.e., before averaging 8 neighboring bins (lowest frequency of 10$^{-4}$ Hz), resulted in $\chi^2$ values of 1082 (996 dof) and 1077 (994 dof) for the former (constant+QPO) and latter (powerlaw+constant+QPO) models, respectively. In summary, the improvement in $\chi^2$ was limited, and thus a powerlaw component is not formally required by the data.

\subsubsection{Kolmogorov-Smirnov and Anderson-Darling goodness of fit tests:}\label{supsec:kstest}
To investigate this even further we also performed the Kolmogorov-Smirnov (K-S) and Anderson-Darling goodness-of-fit tests under the null hypothesis that the noise powers between 0.001 and 0.1 Hz, excluding the three bins of the QPO candidate, are $\chi^2$ distributed with 2$\times$8$\times$8 dof scaled by a factor of 1/8$\times$8\cite{vanderklis88}. In other words, the null hypothesis is that the underlying noise is white. We choose an upper limit of 0.1 Hz to make sure the sample is not biased by higher ($>$0.1 Hz) frequencies.  %The two factors of 8 are because we averaged in frequency by a factor of 8 and averaged 8 individual power spectra. 

\begin{figure}[ht]
 \begin{center}
 %\vspace{-0.75cm}
 \includegraphics[width=6.25in, angle=0]{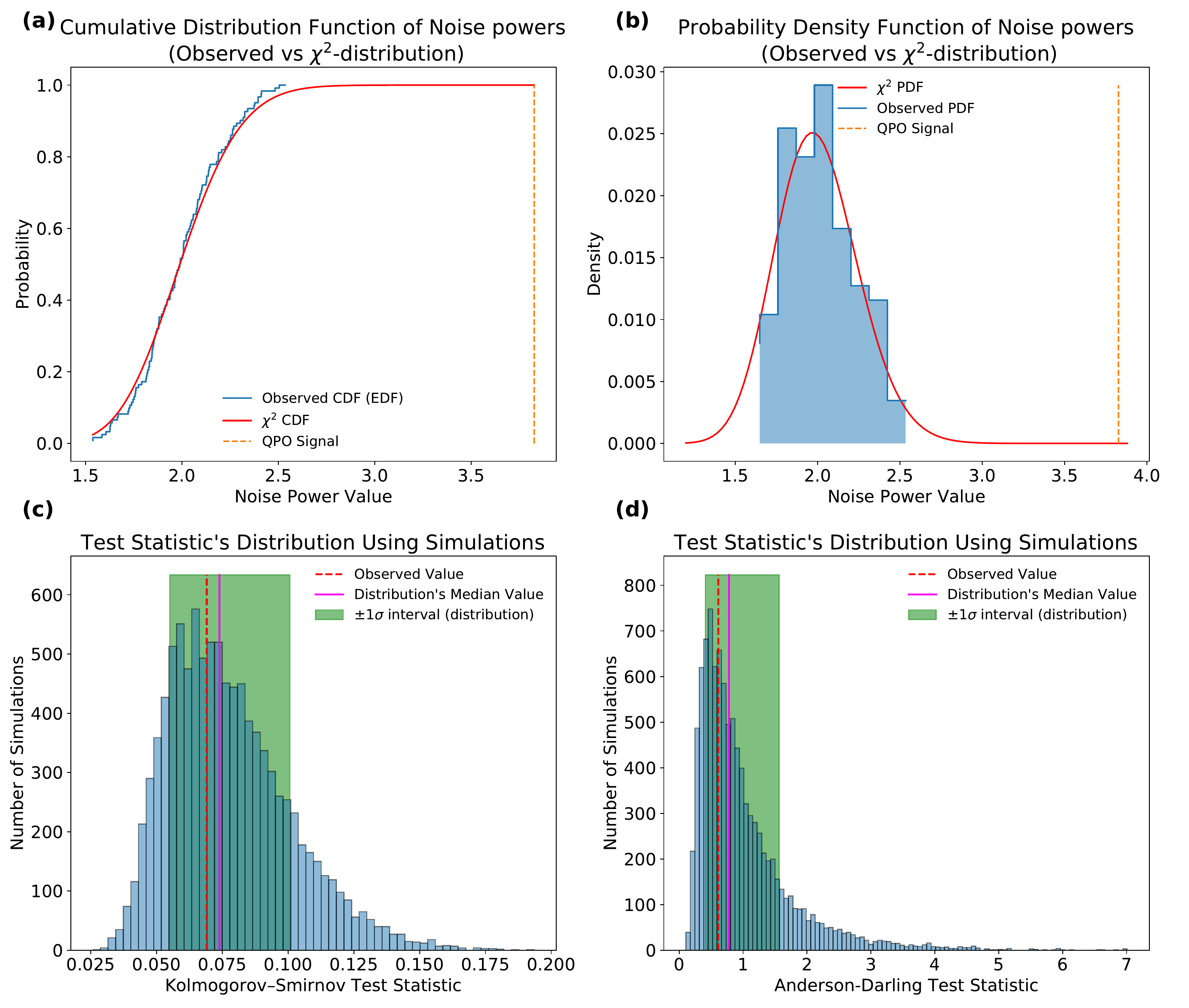}
 \end{center}
 %\vspace{-.35cm} 
 \caption{{\bf Tests for white noise of ASASSN-14li's PDS.} {\bf (a)} The empirical distribution function (EDF) of ASASSN-14li's noise powers in 0.001-0.1 Hz, i.e., surrounding the QPO signal at 7.65 mHz. The blue histogram is the data while the red curve is the expected $\chi^2$ distribution for white noise. The EDF tracks the expected CDF quite well over the range of observed power values. The dashed orange line marks the power value of the highest bin in the QPO. {\bf (b)}. Here we compare the probability density function of the observed noise powers with the expected $\chi^2$ distribution for white noise. {\bf (c)} Distribution of the K-S test statistic for a sample size of 122 (=number of bins between 0.001 and 0.1 Hz minus the three QPO bins) for a $\chi^2$ distribution. The observed test statistic value (dashed red line) lies very close to the median of the distribution (magenta line) and is thus consistent with ASASSN-14li's noise powers being white (see sec. \ref{supsec:kstest}). {\bf (d)} Distribution of Anderson-Darling test statistic for a sample size of 122 for a $\chi^2$ distribution. Again, it is evident that ASASSN-14li's noise powers are consistent with the expected $\chi^2$ distribution.} %The red region is again the $\pm$1-$\sigma$ uncertainty on the observed value.}The red region is the $\pm$1-$\sigma$ uncertainty on the observed test statistic value considering the error bars on the individual powers in the observed PDS (see Fig. \ref{fig:xmmchan_swift_pds})
 \label{fig:whitenoisetests}
 \end{figure}

Before evaluating the K-S and Anderson Darling tests, we computed the empirical distribution function (EDF) and the probability density function (PDF) of the noise powers. These are shown in the top panels of Fig. \ref{fig:whitenoisetests} (blue histograms) along with the expected $\chi^2$ distribution (red). Again, it can be seen that the data track the expected $\chi^2$ distribution quite well. 

We then computed the K-S test statistic using the EDF. Thereafter, we generated bootstrap simulations to compute the distribution of the K-S test statistic itself as follows. The distribution of the test statistic is necessary to reject or not to reject the null hypothesis. First, we randomly draw the same number (=122) of elements as the observed noise powers from a $\chi^2$ distribution with 128 dof. Then we evaluate its EDF and scale it by a factor of 1/64, just like the real data. Finally, we compute the K-S test statistic and store its value. We repeat this process 10,000 times to get a distribution of the K-S test statistic for a $\chi^2$ distribution with 128 dof for a given sample size of N$_{sample}$ (=122). It is important to realize that this bootstrap method properly accounts for the size of the sample. The resulting distribution is shown as a blue histogram in the bottom left panel of Fig. \ref{fig:whitenoisetests}. ASASSN-14li's observed K-S test statistic is indicated by a dashed vertical red line and is close to the median value of the distribution (magenta vertical line). This suggests that noise powers are very much consistent with the expected $\chi^2$ distribution, i.e., ASASSN-14li's noise powers in the vicinity of the QPO candidate signal are consistent with being white. %Assuming a critical value of 0.05, the null hypothesis that ASASSN-14li's noise powers are $\chi^2$ distributed cannot be rejected. 

To further test the null hypothesis we also investigated the goodness-of-fit with the Anderson-Darling test. We computed the distribution of the test statistic using the same bootstrap technique described above. This resulting distribution is shown in the bottom right panel of Fig. \ref{fig:whitenoisetests}. Again, it is evident from the value of Anderson-Darling's test statistic (dashed red line) that the observed noise powers of ASASSN-14li's PDS are consistent with being $\chi^2$ distributed. 

We also repeated the above tests by changing the frequency upper limit from 0.1 to 0.2 and 0.5. All of them lead to the same conclusion that the noise powers above 0.001 Hz are consistent with being white.

\subsubsection{Statistical Significance under white noise:}
% Both the Kolmogorov-Smirnov and Anderson-Darling tests suggest that the noise powers in ASASSN-14li between 0.001 and 0.5 Hz are consistent with being $\chi^2$ distributed, i.e., consistent with white noise.
We estimate the statistical significance of the feature at 7.65 mHz under the white noise hypothesis as follows. First, we ensured that the mean noise level was equal to 2 as this is the value expected from pure Poisson (white noise) process. We then computed the probability, at the 99.73\% (3$\sigma$), the 99.9937\% (4$\sigma$), and the 99.9999426697\% (or 5$\sigma$) confidence levels, of obtaining the power, P = P$_{*}\times$8$\times$8 from a $\chi^2$ distribution with 2$\times$8$\times$8 degrees of freedom. Here P$_{*}$ is the power value of a statistical fluctuation at a given confidence level. As mentioned above this $\chi^2$ distribution was used because we averaged in frequency by a factor of 8 and averaged 8 individual power spectra. Considering all the trials below 0.5 Hz we computed the 3$\sigma$ (1/(371$\times$trials)), the 4$\sigma$ (1/(15787$\times$trials)) and the 5$\sigma$ (1/(1744278$\times$trials)) confidence contours. These are shown as the blue, magenta and the red contours, respectively, in the left panel of Fig. \ref{fig:xmmchan_swift_pds}. The highest bin in the feature at 7.65 mHz is significant at the 4.8$\sigma$ confidence level.

\subsection{Statistical Significance under red noise:}\label{supsec:rednoise}
The analysis above suggests that ASASSN-14li's observed power spectrum within the frequency range of 0.001-0.5 Hz is consistent with white noise. However, it is still plausible that a weak/unknown red noise component is present in the data. To estimate the strength of the red-noise component, we fit the unbinned PDS with a power-law + constant + QPO model. We choose the unbinned PDS as it allows us to access information at frequencies as low as 10$^{-4}$ Hz. Even then, the best-fit power-law normalization and index values were only poorly constrained to be (1.2$\pm$4.5)$\times$10$^{-4}$ and -1$\pm$0.4, respectively. To test the significance of the QPO under red noise in a more systematic manner we estimated the best-fit normalization of the power-law component by fixing the index at various values between -0.3 and -1.7 ($\approx$95\% confidence interval). The resulting normalization values are listed in Table. \ref{table:powlawnoise}. 

\begin{figure}[ht]
 \begin{center}
  \vspace{-1cm}
 \includegraphics[width=6.5in, angle=0]{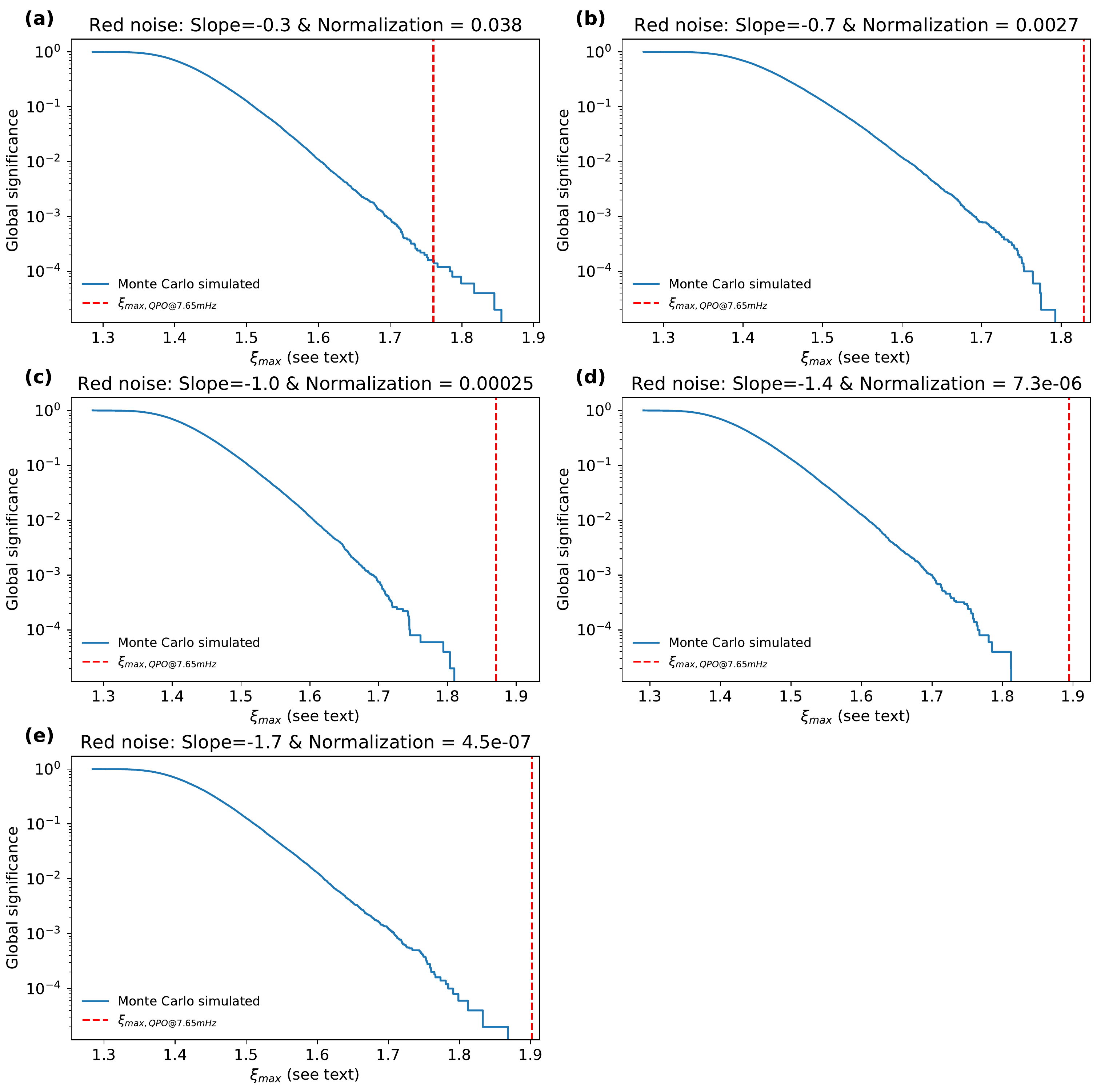}
 \end{center}
 \vspace{-.35cm} 
 \caption{{\bf Global statistical significance under red noise} {\bf (a)} Results from Monte Carlo simulations showing the global probability to exceed (statistical significance) the maximum normalized power below 0.5 Hz ($\xi_{\rm max}$) under the assumption that the red noise has a power-law index of -0.3 and a normalization of 0.038. The specific value of 0.038 is the 1$\sigma$ upper limit on the best-fit normalization for a fixed index of -0.3. Therefore, the statistical significance should be considered conservative, i.e., a lower limit (see sec. \ref{supsec:rednoise}). The normalized power value in the highest QPO bin is indicated by the dashed red line. {\bf (b), (c), (d), (e)} Same as {\bf (a)} but for a red noise index of -0.7, -1.0, -1.4, and -1.7, respectively. Again, their respective normalization values are 1$\sigma$ upper limits estimated directly from modeling ASASSN-14li's unbinned PDS (see sec. \ref{supsec:modelpsd}).}
 \label{fig:rednoisesigs}
 \end{figure}

For each power-law index ($\alpha_0$ in -0.3, -0.7, -1.0, -1.4, -1.7) and its corresponding upper limit on the normalization, i.e., best-fit normalization + uncertainty on the best-fit normalization ($N_{*}$; see Table. \ref{table:powlawnoise}), we employed the following rigorous Monte Carlo approach to estimate the {\it global} statistical significance of the QPO:

\begin{enumerate}
        \item Using the algorithm described by Timmer \& Koenig (1995)\cite{timmerkonig95} we simulated 8$\times$50,000 Leahy-normalized red noise light curves whose PSD is defined by ($\alpha_0$, $N_{*}$). Each of the light curves were 500,000 s in length, i.e., a factor of 50 longer than the length of the light curves used in ASASSN-14li's PSD (Fig. \ref{fig:xmmchan_swift_pds}). From each of these 8$\times$50,000 simulated light curves we extracted 10,000 s segments from the center in order to account for red noise leakage\cite{uttley02}. The light curves were sampled with a resolution of 1-s similar to ASASSN-14li's data. Leahy power spectra were extracted and sets of 8 PDS were combined to obtain an average PDS. Finally, we averaged 8 neighboring bins in each of the 50,000 power spectra obtained from averaging 8 individual PDS. This gave us 50,000 simulated power spectra described by a given red-noise and that are sampled and averaged exactly as real data. The light curves and consequently the average PDS were fairly computationally expensive to simulate. To simulate 50,000 PDS required a total computation time of $\approx$5 hours on a 64 core machine. We used {\tt Amazon web services} for multiprocessing.  
        
        \item In order to carry out a global search for a signal below 0.5 Hz, we employed a methodology similar to Benlloch (2001)\cite{benlloch01}. We first divide each of the 50,000 simulated PDS (P$_{\rm i}$, i = 0, 1 ... 49999) with the average of all the 50,000 simulated power spectra ($\langle$ P$_{\rm i}$ $\rangle$). Then for each of the normalized (simulated) PDS we note down the maximum power spectral feature below 0.5 Hz, $\xi_{\rm max}$ = max$\Big[$P$_{\rm i}$/$\langle$ P$_{\rm i}$ $\rangle\Big]$. This way we emulate a ``global'' search that includes all frequency bins (trials) below 0.5 Hz and also properly accounts for red-noise. 
        
        \item We then estimate $\xi_{\rm max}$ for the observed data in Fig. \ref{fig:xmmchan_swift_pds}, $\xi_{\rm max, obs}$, by simply dividing the observed PDS by $\langle$ P$_{\rm i}$ $\rangle$.
        
        \item Finally, using the 50,000 values of $\xi_{\rm max}$ we computed (1.0 minus the cumulative distribution function) plot, i.e., the probability to exceed a given $\xi_{\rm max}$ value, and compare that with the observed QPO's value, i.e., $\xi_{\rm max, obs}$. The results are shown in Fig. \ref{fig:rednoisesigs}. 
        
   \end{enumerate}     
It is evident that for all the cases, the observed QPO value is significant at greater than at least 10$^{-4}$ or the 3.9$\sigma$ level. The reported significance values should be considered as the lower limits as they correspond to the upper limits of best-fit red-noise normalizations (see Table \ref{table:powlawnoise}).
   
%++++++++++++++++++++++++++++++++++++++++++++++++++++++++++++++++++++++++++++++++++++++++++++++++++++++++++++++++++++++++++++++++++

\subsection{Separate {\it XMM-Newton} and {\it Chandra} PDS.}
\begin{figure}[ht]
 \begin{center}
 \vspace{-0.75cm}
 \includegraphics[width=5.5in, angle=0]{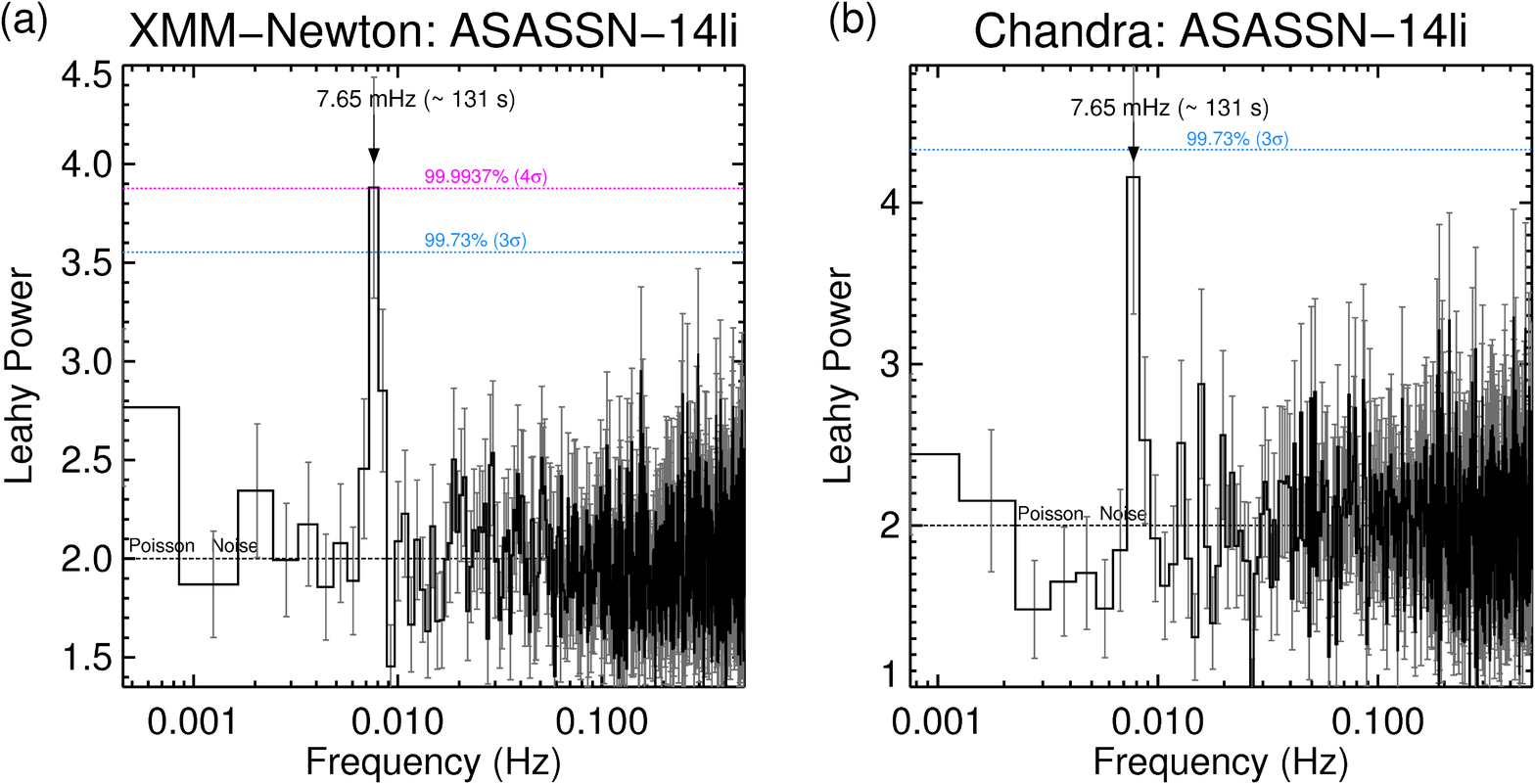}
 \end{center}
 \vspace{-.35cm} 
 \caption{{\bf {\it XMM-Newton} and {\it Chandra} power spectra of ASASSN-14li and 
 backgrounds.} {\bf (a)} ASASSN-14li's average x-ray (0.3-1.0 keV) power density 
 spectrum using six continuous 10,000 s light curves taken with {\it XMM-Newton}. The 
 frequency resolution is 0.8 mHz. The strongest feature in the power spectrum lies at 
 a frequency of 7.65$\pm$0.4 mHz ($\approx$ 131 seconds). The dashed horizontal lines 
 represent the 3 and the 4$\sigma$ global statistical contours, i.e., takes into 
 account a search at all frequencies below 0.5 Hz. $\pm$1$\sigma$ error bars are overlaid as grey bars. {\bf (b)} ASASSN-14li's x-ray (0.4-1.0 keV) power 
 density spectrum using twelve 2,000 s light curves taken with {\it Chandra}'s ACIS 
 instrument. The frequency resolution is 1 mHz. The strongest feature again lies at 
7.75$\pm$0.5 mHz and is consistent with the most prominent feature in the average 
{\it XMM-Newton} spectrum (see {\bf (a)}). 
% {\bf (c, d)} The PDS of the background 
% light curves taken with {\it XMM-Newton} and {\it Chandra}/ACIS. The PDS were evaluated 
% in the same manner as their respective source power spectra (top panels), i.e., same 
% light curve length and frequency resolution. The vertical green band shows the location 
% of the 7.65 mHz QPO and is clearly absent in both the {\it XMM-Newton} and the {\it Chandra} 
% background light curves. 
}
 \label{fig:separate_pds}
 \end{figure}

After establishing the QPO at 7.65 mHz we extracted an average PDS
separately from {\it XMM-Newton} and {\it Chandra} data. These are 
shown in the top-left and the top-right panels of Fig. \ref{fig:separate_pds}. 
The 7.65 mHz QPO is evident in both the detectors and is significant at the $\approx$4$\sigma$ and $\gtrsim$ 2.6$\sigma$ levels assuming a global search 
including all trials below 0.5 Hz. The fact that the QPO is present in two different detectors at 
different epochs is ensuring that the signal is detector-independent, 
albeit it is not highly statistically significant in the {\it Chandra} data by itself.

%++++++++++++++++++++++++++++++++++++++++++++++++++++++++++++++++++++++++++++++++++++++++++++++++++++++++++++++++++++++++++++++++++

\subsection{Stacked {\it Swift}/XRT PDS.}
The XRT on board {\it Swift} has an effective area\footnote{https://swift.gsfc.nasa.gov/about\_swift/xrt\_desc.html} 
of $\lesssim$ 1/20$^{th}$ that of {\it XMM-Newton}/EPIC's combined pn+MOS\footnote{https://xmm-tools.cosmos.esa.int/external/xmm\_user\_support/documentation/uhb/epicfilters.html}. Nevertheless, using 1000-second light curve segments spread across
the $\gtrsim$ 450 d flare we constructed an average 0.3-1.0 keV PDS. Similar to {\it XMM-Newton} extraction, in order to mitigate the pile-up issue but at the same time not compromise on the count rate, we considered only grade 0 events within an 25'' circular region. The QPO at 7.65 mHz is recovered at over the 3$\sigma$ (single trial) 
confidence level. This is shown in the right panel of Fig. \ref{fig:xmmchan_swift_pds}. 
This again suggests that the QPO was stable over the first 450 d since its 
discovery.

%++++++++++++++++++++++++++++++++++++++++++++++++++++++++++++++++++++++++++++++++++++++++++++++++++++++++++++++++++++++++++++++++++

\subsection{The 7.65 mHz QPO is stable.}
The fact that the 7.65 mHz QPO is present in the average PDS of eight
observations scattered over 450 d already demonstrates that the QPO is 
stable throughout the first 450 d of the flare. To establish this 
further we constructed a dynamic PDS where we show the progress of the 
PDS as we add one additional PDS (see supplement movie). This demonstrates 
clearly that the 7.65 mHz QPO does not originate from a single observation 
that dominates the average PDS. Instead, the signal gradually improves as 
more and more data is added. This implies that the signal is present to some
extent in all the individual {\it XMM-Newton} and {\it Chandra} power spectra. 
Furthermore, the average {\it Swift} PDS taken over 450 d also indicates that 
the QPO has to be stable.

%++++++++++++++++++++++++++++++++++++++++++++++++++++++++++++++++++++++++++++++++++++++++++++++++++++++++++++++++++++++++++++++++++

\section{\Large{\bf QPO's Coherence, fractional rms amplitude, and folded light curves.}}\label{supsec:coherms}
The coherence of the QPO (centroid frequency ($\nu$)/width ($\Delta\nu$)) in the combined {\it XMM-Newton} and {\it Chandra} power spectrum can be estimated from the unbinned power spectrum obtained by averaging the 8 individual PDS, i.e., before averaging neighboring frequency bins. Modeling the QPO with a Lorentzian functional form results in a best-fit centroid and width of 7.89$\pm$0.1 mHz and 0.5$\pm$0.2 mHz, respectively. Combining these two values, the coherence of the QPO is 16$\pm$6.

To visualize this modulation in the time domain, we first folded the {\it Chandra} (C1) light curve as it is not severely effected by pile-up (see sec. \ref{supsec:chandrapileup}). We estimated the fold period from the power spectrum of the light curve over-sampled by a factor of 3, i.e., total light curve length is 3 times the original C1 light curve. Oversampling translates to subtracting the mean from the light curve and padding the end with zeros and then computing the power spectrum. Oversampling is a commonly used technique in pulsar period searches\cite{deeptothesis, middleditchthesis, ransom}. For folding purposes, we used the frequency that corresponds to the highest power within 7.65$\pm$0.7 mHz, where 0.7 mHz is the upper limit on the QPO's width (see above). For C1 this period is 134.6$\pm$0.1 seconds (or 7.43$\pm$0.006 mHz). The error on the period was estimated using Eq. 3.12 of Chakrabarty (1999)\cite{deeptothesis} which was derived by Middleditch (1976)\cite{middleditchthesis}. The resulting folded light curve is shown in Fig. \ref{fig:foldedchandra}. The best fit sinusoidal curve is overlaid in red. The fractional modulation amplitude derived from the best-fit {\bf sinusoidal curve is 35$\pm$8\%} and is consistent {\bf (within the 90\% confidence limit)} with the measurement from the power spectrum which yields 59$\pm$11\%. 

\begin{figure}[ht]
 \begin{center}
 \vspace{-0.75cm}
 \includegraphics[width=5.in, angle=0]{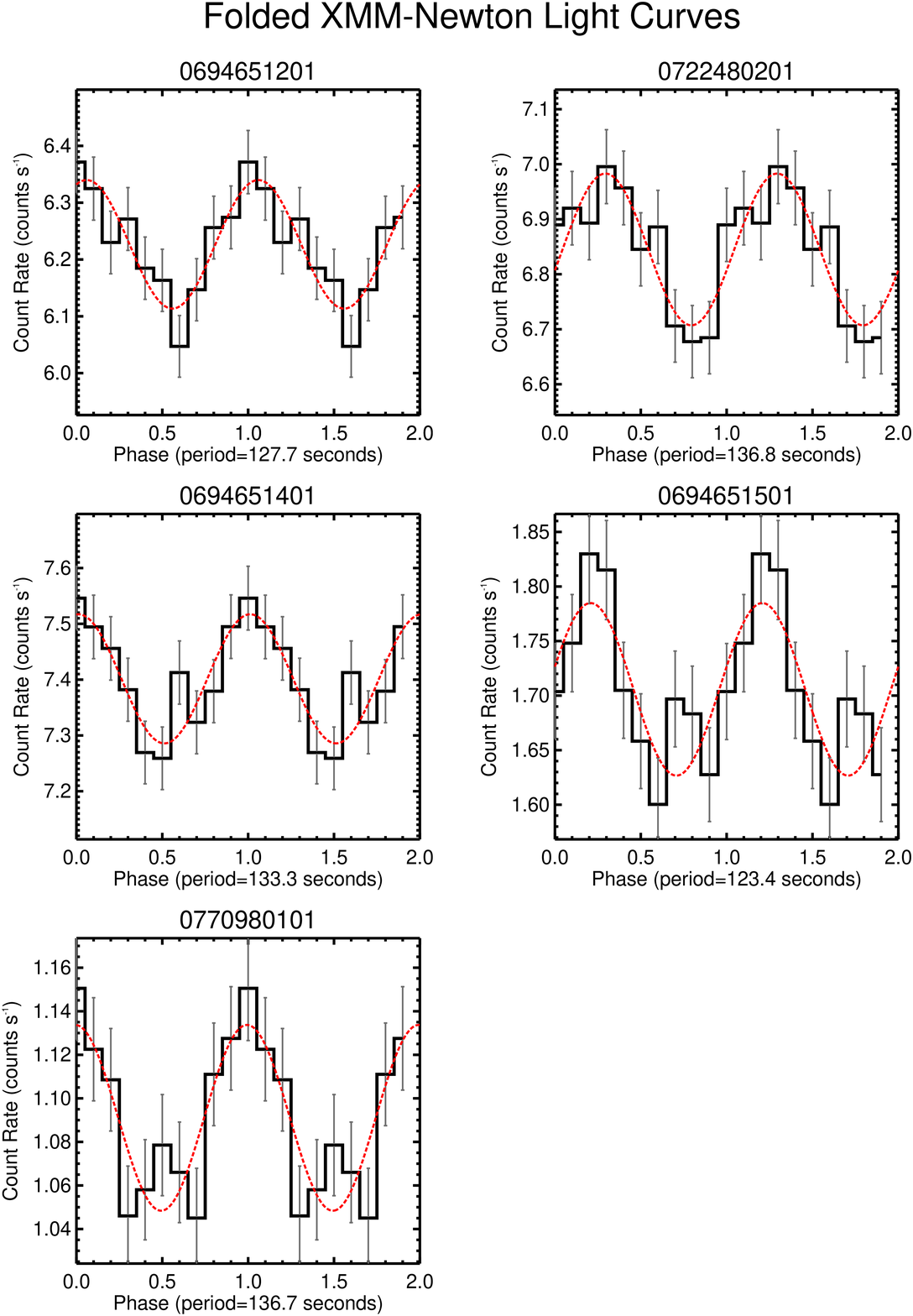}
 \end{center}
 \vspace{-.35cm} 
 \caption{{\bf Folded {\it XMM-Newton} light curves.} Same as Fig. \ref{fig:foldedchandra}. In each case, the fold period corresponds 
 to the frequency of the highest power spectral peak between 7.65$\pm$0.7 mHz in the PDS of the 
 longest GTI oversampled by a factor of 3 (see Table. \ref{table:tableqporms}). Except for 0770980101 the rest of 
 the datasets suffered from pile-up. Therefore, their respective fractional rms amplitudes should 
 be treated as lower limits.}
 \label{fig:xmmfolds}
 \end{figure}

{\bf {\it Chandra} observations are typically dithered and as a result the source region does not lie on the same pixels throughout the exposure. The detector has some bad columns/pixels and if the source region dithers in and out of these bad pixels this can alter the true count rate. Because the nominal periods for ACIS in the two dither directions are 1000 and 707 s, it is unlikely to produce any periodic modulation on a timescale of 131 seconds. However, it is plausible that the spacecraft dithering could effect the amplitude of the QPO. To investigate this possibility we used the {\it Chandra} {\tt ciao} tool {\tt dither\_region} to estimate the fractional area of the source region as a function of time during the 25 ks exposure. The source area fraction was unity throughout the observation and therefore we conclude that dithering, and hence bad pixels, could not have affected the rms of the QPO during epoch C1.

We also folded the {\it XMM-Newton} observations at their respective fold periods. Similar to C1 above the fold period for each {\it XMM-Newton} observation corresponds to the frequency of the highest power bin within 7.65$\pm$0.7 mHz in the oversampled PDS. The resulting folded light curves are shown in Fig. \ref{fig:xmmfolds} and are described in Table \ref{table:tableqporms}. The power spectra from individual {\it XMM-Newton} observations do not have enough signal-to-noise to resolve the QPO and therefore it was not possible to estimate the QPO rms directly from the PDS. Instead, we estimated the fractional rms by fitting a sinusoidal curve to the folded light curves (see Table. \ref{table:tableqporms}). Except for epoch X5 the rest of the {\it XMM-Newton} observations are piled-up, and therefore the rms amplitudes should be treated as the lower limits. However, because of low level of pile-up in X5 we think the $\approx$3\% value is close to the true fractional rms amplitude of the QPO during at least epoch X5. 
}

\begin{figure}[ht]
 \begin{center}
 \vspace{-0.75cm}
 \includegraphics[width=4in, angle=0]{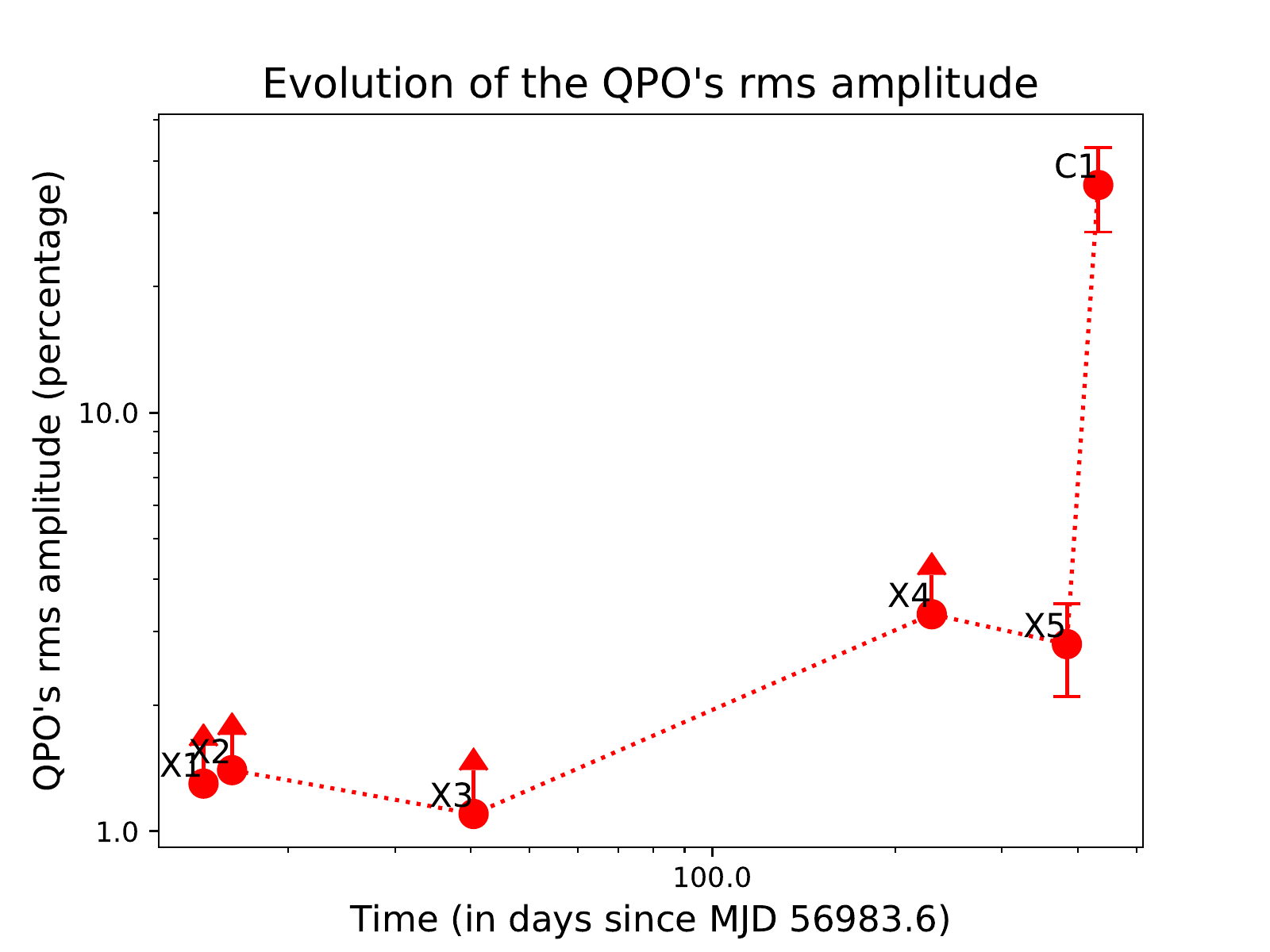}
 \end{center}
 %\vspace{-.35cm} 
 \caption{{\bf QPO's rms evolution.} X1, X2, X3 and X4 data were piled-up and hence their rms amplitudes are shown as lower limits. A sharp rise in the rms between X5 and C1 is evident (see Table \ref{table:tableqporms}).}
 \label{fig:rmsvstime}
 \end{figure}

The fact that this value is significantly lower than the rms during C1 suggests that the QPO underwent a strong amplification between X5 and C1 which were separated by a duration of roughly 50 days. {\bf The long-term evolution of the QPO's fractional rms amplitude is shown in Fig. \ref{fig:rmsvstime}.}

%++++++++++++++++++++++++++++++++++++++++++++++++++++++++++++++++++++++++++++++++++++++++++++++++++++++++++++++++++++++++++++++++++

\section{\Large{\bf Estimating {\it Chandra}/ACIS Pile-up Fraction.}}\label{supsec:chandrapileup}
The mean count rate during the {\it Chandra} observation (C1 in Fig. \ref{fig:swiftlc}) was only roughly 0.008 counts/sec. At such low count rates pile-up is expected to be minimal. Nevertheless, we estimated the pile-up fraction using the {\tt ciao} tool {\tt pileup\_map}. Using the counts/frame in the brightest pixel in the pile-up image generated from this tool we calculated the pile-up fraction to be only $\approx$ 4.5\% (see Eq. 3 of pile-up analysis guide\footnote{http://cxc.harvard.edu/csc/memos/files/Davis\_pileup.pdf}). Using PIMMS\footnote{http://cxc.harvard.edu/toolkit/pimms.jsp} also gives a similar value. In summary, because the {\it Chandra} data were not severely piled-up we could estimate the rms value of the QPO during the C1 epoch in Fig. \ref{fig:swiftlc}. 

%++++++++++++++++++++++++++++++++++++++++++++++++++++++++++++++++++++++++++++++++++++++++++++++++++++++++++++++++++++++++++++++++++

\section{\Large{\bf Ruling out a Pulsar Origin.}}
A pulsar origin for ASASSN-14li--and thus the QPO--is unlikely for many reasons. 
\begin{enumerate}
\item Firstly, the size of ASASSN-14li's optical/UV photosphere ($\sim$ 10$^{14}$ cms\cite{pasham17,brownjs17}) is a factor of $\gtrsim$10$^{5}$ larger than the characteristic emission size of a 2M$_{\odot}$ neutron star's accretion disk of a few thousand gravitational radii. A stellar-mass BH origin can also be ruled out on the same basis. 
\item Secondly, ASASSN-14li's radio emission is not dominated by emission from a neutron star. Its radio spectral energy distributions are consistent with synchrotron self-absorption with a characteristic emission size of a few$\times$10$^{16}$ cms\cite{alexander16,pashamsjort17}. Again, this is several orders of magnitude large than a typical neutron star's size of roughly 10$^{6}$ cms.
\item ASASSN-14li's host galaxy distance of 90.3 Mpcs\cite{Holoien16} would imply that the putative neutron star is emitting at an apparent bolometric (x-ray+optical+UV) luminosity $>$3$\times$10$^{6}$ its Eddington limit. This is plausible in light of the recent discovery of so-called ultra-luminous x-ray (ULX) pulsars\cite{bachetti14} with maximum luminosities upto 7$\times$10$^{40}$ erg/sec. However, ASASSN-14li would still be an extreme ULX pulsar with a factor of $>$1400 brighter than even the most luminous ULX pulsar known\cite{Israel17}. Moreover, all three known ULX pulsars and the bursting pulsar GRO J1744-28\cite{jahoda99} are highly variable\cite{bachetti14,Israel17}. They reach super-Eddington luminosities only for brief periods of a few days at a time\cite{bachetti14,Israel17}. ASASSN-14li on the other hand has an average apparent bolometric luminosity of $>$5$\times$10$^{43}$ erg/sec or $\approx$2$\times$10$^{5}$ times Eddington for a neutron star for over at least two years after its discovery\cite{brownjs17}. 
\item Also, ASASSN-14li's x-ray spectrum is unlike any ULX pulsar. Because all X-ray bright jets show hard x-rays, if ASASSN-14li's x-ray emission were highly beamed one would expect hard x-rays to be present. This is contrary to the observed very soft x-ray spectrum\cite{pasham17,miller15}. 
\item The observed 7.65 mHz feature has a finite width (Fig. \ref{fig:xmmchan_swift_pds}; coherence=16$\pm$6)) unlike a pulsar's signal that is expected to be highly coherent (see sec. \ref{supsec:coherms} for more discussion).
\item In principle, ASASSN-14li could be a foreground pulsar that happened to spatially coincide with a background galaxy. This would be highly coincidental especially because there are no known pulsars in this sky region\cite{pulsarcatalog}. Nevertheless, we estimated the chance coincidence with a background galaxy as: 
\begin{equation}
\frac{N_{gals} \times \pi R_{x\textnormal{-}ray}^{2}}{\pi R_{gal}^{2}} 
\end{equation}
\noindent where $N_{gals}$ is the number of galaxies within a circle of radius $R_{gal}$ and centered on ASASSN-14li. $R_{x\textnormal{-}ray}$ is the typical positional uncertainty of {\it Chandra}/ACIS which has been estimated to be 0.8'' (90\% positional accuracy\footnote{http://cxc.harvard.edu/cal/ASPECT/celmon/}). Using the galaxy catalog from the Sloan Digital Sky Survey (data release 14) we find $N_{gals}$=1505 within a circular area of radius 10'. This translates to a chance coincidence of less than 3\%. The mean and the lowest $g$-band magnitude of galaxies around ASASSN-14li is 22.9 and 28.4, respectively while ASASSN-14li's host galaxy--prior to the TDE--had a $g$-band magnitude of 16.1\cite{Holoien16}. We repeated this estimate with a sky area of $\pi$5'$\times$5' and $\pi$15'$\times$15' to find that the resulting chance probabilities are the same. We stress that the above 3\% estimate can be considered conservative (upper limit) as it includes chance coincidence with any part of the galaxy not just the center. 
\item Finally, ASASSN-14li's multiwavelength properties are unlike any neutron star outburst and are all similar to many previously known TDEs. A pulsar origin would only then compel us to conclude that all previously-known TDEs are foreground x-ray pulsars that perfectly coincided with background galaxies, which seems even more unlikely.
\end{enumerate}

%++++++++++++++++++++++++++++++++++++++++++++++++++++++++++++++++++++++++++++++++++++++++++++++++++++++++++++++++++++++++++++++++++

 \section{\Large{\bf Description of the Five Frequencies of Motion around a Black Hole.}}
A test particle moving in the strong gravity of a black hole has three fundamental frequencies. The fastest at any given radius is the Keplerian orbital frequency ($\nu_{\phi}$) for motion in the equatorial plane. Perturbations can induce two additional frequencies in the radial and the vertical directions. These are known as the radial ($\nu_{r}$) and the vertical epicyclic ($\nu_{\theta}$) frequencies, respectively. Furthermore, beating between these three coordinate frequencies can lead to two additional frequencies: $\nu_{LT}$ = $\nu_{\phi}$ - $\nu_{\theta}$ and $\nu_{per}$ = $\nu_{\phi}$ - $\nu_{r}$. These are known as the Lense-Thirring precession and the periastron precession frequencies, respectively. The frequencies are defined as follows:

\begin{equation} {\rm 
\nu_{\phi} = \pm \frac{c^3}{2\pi GM}\bigg[ \frac{1}{r^{3/2} \pm a}\bigg]
}
\end{equation}

\noindent where r is the radius in units of gravitational radius, $R_{g}$ = GM/c$^{2}$. $G$, $M$, and $c$ are the gravitational constant, black hole mass, and the speed of light, respectively. $a$ is the black hole's dimensionless spin parameter defined as $a$ = J/(GM/c$^{2}$). $J$ is the black hole's angular momentum.

\begin{equation}
     \nu_{\theta} = \nu_{\phi}\bigg[1 \mp \frac{4a}{r^{3/2}} + \frac{3a^2}{r^2} \bigg]^{1/2}
 \end{equation}
 
 \begin{equation}
     \nu_{\rm r} = \nu_{\phi}\bigg[ 1 - \frac{6}{r} \pm \frac{8a}{r^{3/2}} - \frac{3a^2}{r^2}\bigg]^{1/2}
 \end{equation}
 The upper and lower signs in the above equations refer to the prograde and retrograde orbits, respectively (See Motta et al. (2014)\cite{motta14} and Franchini et al. 2017\cite{franchini17} for more details). The equations are exact results for the Kerr geodesics and were derived by Kato (1990)\cite{kato1990}.
 
 %++++++++++++++++++++++++++++++++++++++++++++++++++++++++++++++++++++++++++++++++++++++++++++++++++++++++++++++++++++++++++++++++++

 \section{\Large{\bf Potential QPO Mechanisms.}}

 In general, because the victim star can approach the disrupting BH from any direction, its orbital plane is expected to be arbitrarily oriented with respect to the BH's spin axis. The transient accretion disk that forms after disruption is thus expected to be born significantly misaligned with respect to the BH’s spin axis, in contrast to most accreting BHs (which, as long-lived systems, are expected to exhibit spin-orbit alignment). Past work predicted\cite{nick_lens_thir, franchini16_tde} that the aspherical spacetime around a spinning BH should force such a misaligned disk to precess as a roughly rigid body and produce quasi-periodic modulation of the soft x-ray flux, as is seen in many general relativistic magnetohydrodynamic (GRMHD) simulations of tilted thick disks \cite{fragile07, liska17}. Assuming the observed 7.65 mHz QPO originates from the global precession of a newly-formed accretion disk would imply, however, that the precessing disk/ring is very narrow. Using the semi-analytical approach for a precessing TDE disk as formulated in \cite{franchini16_tde}, for a BH mass between 10$^{4}$ and 10$^{7}$ $M_{\odot}$, the implied radial extent of the disk must be between a few tens to a fraction of an R$_{g}$, respectively, even for maximally spinning BHs. Even narrower disks are required for smaller spin values.  %Recent work by Motta et al. \cite{mottafran18} have argued that disks extending beyond a few gravitational radii would soon align with the BH's spin axis, in contrast to the remarkably stable signal seen here. 
 %Furthermore, such systems are yet to be realized in full 3-dimensional general relativistic magneto-hydrodynamic (GRMHD) simulations. 
  
In order to produce such narrow disks, the star would need to plunge deep into the gravitational potential of the BH. The likelihood of this can be quantified with the so-called penetration parameter, $\beta$, \cite{stonesariloeb}, defined as the ratio of the tidal radius (the radius at which the BH's tidal forces exceed the star's internal pressure) and the pericenter radius (distance of the star's closest approach). For ASASSN-14li, if global disk precession is the origin of the QPO, the penetration parameter would have to be very high. Assuming a 10$^{6}$ $M_{\odot}$ BH, $\beta$ has to be $\approx$ 25-50 if we wish to tune the pericenter (and initial disk outer edge) to the ISCO scale for a rapidly spinning Kerr BH. %The required $\beta \sim 100$ increases to a few hundreds for lower masses. 
  
  As the dynamical fraction of TDEs with penetration parameters $>\beta$ is at most $\approx 1/\beta$ \cite{tdebeta2}, it requires significant fine-tuning to produce narrow accretion tori for BHs in the mass range inferred for ASASSN-14li. Moreover, the details of the disk formation process in TDEs are complicated and subject to significant theoretical debate \cite{Shiokawa+15, Hayasaki+17}, although for very relativistic pericenters disk formation may proceed more efficiently \cite{Dai+15}.  Even if we could produce a narrow, efficiently circularized TDE disk for a high-$\beta$ disruption, such a disk would likely expand outward in a quasi-viscous way, although perhaps shocks from returning debris streams could regulate this expansion.  In summary, a global disk precession origin for this QPO seems disfavored by extensive fine-tuning and the observed long-term stability.

Alternatively, GRMHD simulations of tilted accretion flows have shown significant variability from their innermost annuli due to ``plunging streams'' that transport matter from the disk into the BH horizon.  While this variability can occur in the frequency range presented by ASASSN-14li \cite{henisey09}, it exhibits neither the large amplitudes nor the long-term stability we have found \cite{henisey12}, and therefore does not appear to be a promising model for the ASASSN-14li QPO. 

  A third explanation could lie in theoretical work \cite{nixon12,lodatoprice10} that has predicted the occurrence of long-lived, discrete, narrow and nodally precessing rings in the inner regions of a misaligned accretion flows, although the viability of such ``disk tearing'' is still controversial, and has yet to be seen in fully 3-dimensional GRMHD simulations \cite{krolikhawley15}.  It is also unclear why QPOs from these rings would exhibit such remarkable stability over long periods of time, as the global properties of the large-scale TDE disk change dramatically.  
  
One of the clearest qualitative differences between TDE and standard accretion disks is the generic expectation of spin-orbit misalignment. This suggests that the existence of a uniquely high amplitude and stable X-ray QPO in ASASSN-14li may originate in a variability mechanism unique to tilted disks, but none of the three proposed sources of tilted disk variability examined here seem fully satisfactory, posing a significant challenge for theories of TDE disks, and perhaps tilted accretion systems more generally.

Given the evident shortcomings of QPO mechanisms related to spin-orbit misalignment, 
we also consider orbiting hot spots (e.g., \cite{schnittmanhotspot}) as an origin for 
the QPO. The hot spot model suffers from the following three problems:
\begin{enumerate}
    \item Over-dense clumps, which might be the source of the ``hot spots'', tend to shear out
on roughly the local orbital timescale due to differential rotation within the disk.
Thence, it seems unlikely they would result in the fairly high coherence observed
here, unless the duty cycle was quite high.
    \item Assuming hot spots shear out and reform, the narrow range of frequencies seen in
this QPO means the hot spots would necessarily always have to form at the same radius, which 
seems contrived for an isolated disk.  In a TDE disk, however, returning debris streams will 
shock the disk and create hot spots at their point of contact, which is fixed to be the 
pericenter radius of the original star. However, resulting hot spots would orbit the black 
hole too slowly to match the observed QPO frequency, unless the pericenter was quite close 
to the ISCO (see above arguments for the unlikeliness of this). Otherwise, one would expect 
the asymmetry imposed by the gas supply to wash out as the gas flow circularizes into a disk 
(analogous to accretion in low-mass x-ray binaries).
    \item There are instabilities that can produce quasi-stable over-dense clumps in disks (an
example is the Papaloizou-Pringle instability \cite{papalapring84}). However, those ``clumps'' are
manifestations of wave patterns in the flow. Therefore, they move at the pattern
speed \cite{papalapring84}, which is generally slower than the orbital speed \cite{nealon18}, 
exacerbating the problem of matching the very high frequency of this QPO.
% We are then back to the problem of matching the very high frequency of this QPO.
\end{enumerate}

A final possibility for the QPO is that it is associated with the orbital motion of an intact
stellar mass object (a star, stellar core, or compact remnant). This would produce a very 
coherent QPO signal over very long periods. However, this model suffers from the same problem 
as the LT precession--the resulting QPO frequency would be too low to explain the observed 7.65 
mHz signal, unless the black hole were spinning at close to its maximum value and the {\bf perturber} were orbiting near the ISCO. It is unlikely that a stellar mass object could be deposited 
onto a circular orbit by the events leading to the observed TDE--even if a core survived a partial 
disruption, it would not be energetically capable of tidally circularizing.  A hypothetical binary 
companion to the {\bf disrupted} star would likely have been ejected by the Hills mechanism.  If a stellar 
mass object exists on an orbit between the ISCO and the tidal radius, it must predate the tidal 
disruption that triggered ASASSN-14li. The extremely short gravitational wave inspiral time {\bf within a few gravitational radii of 
the ISCO ($t_{\rm GW} \approx 0.4-20$ yr, for orbital frequencies $\Omega=7.65~{\rm mHz}$, black hole masses of $10^{6-7}M_\odot$ and companions between $1-10 M_\odot$)} makes it unlikely that a 
compact object would be in residence there, and a main sequence star could not survive the local 
tidal shear.  The residence times at the tidal radius are roughly $\sim 10^4$ times longer, and 
conservative mass transfer due to Roche lobe overflow (RLOF) can stabilize main sequence stars in 
such orbits for up to $\sim 10^7$ yr \cite{dai13, metzgerstone17}, but the orbital frequency 
there is $\sim 0.1$ mHz, far less than the observed QPO frequency.  {\bf A white dwarf star undergoing RLOF may be the most promising version of this scenario, as its small mass would, for a fixed orbital frequency and black hole mass, produce longer gravitational wave residence times.  If we consider a very low-mass white dwarf, with $M_{\rm WD} = 0.1 M_\odot$, its gravitational wave inspiral time would be $t_{\rm GW} \approx 190~{\rm yr}$ if it were in an $\Omega = 7.65~{\rm mHz}$ orbit around a $10^6 M_\odot$ black hole.  Although this inspiral time is longer than that for other realistic perturbers, it still implies an unacceptably high rate of white dwarf inspirals into supermassive black holes.}

{\bf A final difficulty in explaining this extraordinary QPO is its increasing rms. Of the various QPO models proposed, only the orbiting compact object one seems to present a reasonable explanation: If the TDE disk and compact orbiter are initially in different orbital planes, then as they slowly settle into the same plane, their interactions, and hence the QPO, would strengthen over time. For the other QPO models -- Lense-Thirring precession and orbiting hot spots -- the increasing rms would require that somehow the coherence of the oscillator improves with time. Since effective viscosity is likely to produce dynamical spreading and shearing, it seems more likely that the coherence would decrease, rather than increase, for these models.}

%%%%%%%%%%%%%%%%%%%%%%%%%%%%%%%%%%%%%%%%%%%%%%%%%%%%%%%%%%%%%%%%%%%%%%%%%%%%%%%%%%%%%%%%%%%%%%
% ---- Figure--- Figure ---- Figure--- Figure ---- Figure--- Figure --- Figure--- Figure ----%
%%%%%%%%%%%%%%%%%%%%%%%%%%%%%%%%%%%%%%%%%%%%%%%%%%%%%%%%%%%%%%%%%%%%%%%%%%%%%%%%%%%%%%%%%%%%%%

  \newpage
  \vfill\eject

%%%%%%%%%%%%%%%%%%%%%%%%%%%%%%%%%%%%%%%%%%%%%%%%%%%%%%%%%%%%%%%%%%%%%%%%%%%%%%%%%%%%%%%%%%%%%%
% ---- Figure--- Figure ---- Figure--- Figure ---- Figure--- Figure --- Figure--- Figure ----%
%%%%%%%%%%%%%%%%%%%%%%%%%%%%%%%%%%%%%%%%%%%%%%%%%%%%%%%%%%%%%%%%%%%%%%%%%%%%%%%%%%%%%%%%%%%%%%

\newpage
\vfill\eject

%%%%%%%%%%%%%%%%%%%%%%%%%%%%%%%%%%%%%%%%%%%%%%%%%%%%%%%%%%%%%%%%%%%%%%%%%%%%%%%%%%%%%%%%%%%%%%
% ---- Figure--- Figure ---- Figure--- Figure ---- Figure--- Figure --- Figure--- Figure ----%
%%%%%%%%%%%%%%%%%%%%%%%%%%%%%%%%%%%%%%%%%%%%%%%%%%%%%%%%%%%%%%%%%%%%%%%%%%%%%%%%%%%%%%%%%%%%%%
\newpage
\vfill\eject

%%%%%%%%%%%%%%%%%%%%%%%%%%%%%%%%%%%%%%%%%%%%%%%%%%%%%%%%%%%%%%%%%%%%%%%%%%%%%%%%%%%%%%%%%%%%%%
% ---- Figure--- Figure ---- Figure--- Figure ---- Figure--- Figure --- Figure--- Figure ----%
%%%%%%%%%%%%%%%%%%%%%%%%%%%%%%%%%%%%%%%%%%%%%%%%%%%%%%%%%%%%%%%%%%%%%%%%%%%%%%%%%%%%%%%%%%%%%%

\newpage
\vfill\eject

%%%%%%%%%%%%%%%%%%%%%%%%%%%%%%%%%%%%%%%%%%%%%%%%%%%%%%%%%%%%%%%%%%%%%%%%%%%%%%%%%%%%%%%%%%%%%%
% ---- Figure--- Figure ---- Figure--- Figure ---- Figure--- Figure --- Figure--- Figure ----%
%%%%%%%%%%%%%%%%%%%%%%%%%%%%%%%%%%%%%%%%%%%%%%%%%%%%%%%%%%%%%%%%%%%%%%%%%%%%%%%%%%%%%%%%%%%%%%

 \newpage
 \vfill\eject

%%%%%%%%%%%%%%%%%%%%%%%%%%%%%%%%%%%%%%%%%%%%%%%%%%%%%%%%%%%%%%%%%%%%%%%%%%%%%%%%%%%%%%%%%%%%%%
% ---- Figure--- Figure ---- Figure--- Figure ---- Figure--- Figure --- Figure--- Figure ----%
%%%%%%%%%%%%%%%%%%%%%%%%%%%%%%%%%%%%%%%%%%%%%%%%%%%%%%%%%%%%%%%%%%%%%%%%%%%%%%%%%%%%%%%%%%%%%%

  \newpage
  \vfill\eject

%%%%%%%%%%%%%%%%%%%%%%%%%%%%%%%%%%%%%%%%%%%%%%%%%%%%%%%%%%%%%%%%%%%%%%%%%%%%%%%%%%%%%%%%%%%%%%
% ---- Figure--- Figure ---- Figure--- Figure ---- Figure--- Figure --- Figure--- Figure ----%
%%%%%%%%%%%%%%%%%%%%%%%%%%%%%%%%%%%%%%%%%%%%%%%%%%%%%%%%%%%%%%%%%%%%%%%%%%%%%%%%%%%%%%%%%%%%%%

  \newpage
  \vfill\eject

%%%%%%%%%%%%%%%%%%%%%%%%%%%%%%%%%%%%%%%%%%%%%%%%%%%%%%%%%%%%%%%%%%%%%%%%%%%%%%%%%%%%%%%%%%%%%%
% ---- Figure--- Figure ---- Figure--- Figure ---- Figure--- Figure --- Figure--- Figure ----%
%%%%%%%%%%%%%%%%%%%%%%%%%%%%%%%%%%%%%%%%%%%%%%%%%%%%%%%%%%%%%%%%%%%%%%%%%%%%%%%%%%%%%%%%%%%%%%

  \newpage
  \vfill\eject
  
%%%%%%%%%%%%%%%%%%%%%%%%%%%%%%%%%%%%%%%%%%%%%%%%%%%%%%%%%%%%%%%%%%%%%%%%%%%%%%%%%%%%%%%%%%%%%%
% ---- Figure--- Figure ---- Figure--- Figure ---- Figure--- Figure --- Figure--- Figure ----%
%%%%%%%%%%%%%%%%%%%%%%%%%%%%%%%%%%%%%%%%%%%%%%%%%%%%%%%%%%%%%%%%%%%%%%%%%%%%%%%%%%%%%%%%%%%%%%

 \newpage
 \vfill\eject

%%%%%%%%%%%%%%%%%%%%%%%%%%%%%%%%%%%%%%%%%%%%%%%%%%%%%%%%%%%%%%%%%%%%%%%%%%%%%%%%%%%%%%%%%%%%%%
% ---- Figure--- Figure ---- Figure--- Figure ---- Figure--- Figure --- Figure--- Figure ----%
%%%%%%%%%%%%%%%%%%%%%%%%%%%%%%%%%%%%%%%%%%%%%%%%%%%%%%%%%%%%%%%%%%%%%%%%%%%%%%%%%%%%%%%%%%%%%%

 \newpage
 \vfill\eject

%%%%%%%%%%%%%%%%%%%%%%%%%%%%%%%%%%%%%%%%%%%%%%%%%%%%%%%%%%%%%%%%%%%%%%%%%%%%%%%%%%%%%%%%%%%%%%
% ---- Figure--- Figure ---- Figure--- Figure ---- Figure--- Figure --- Figure--- Figure ----%
%%%%%%%%%%%%%%%%%%%%%%%%%%%%%%%%%%%%%%%%%%%%%%%%%%%%%%%%%%%%%%%%%%%%%%%%%%%%%%%%%%%%%%%%%%%%%%

 \newpage
 \vfill\eject

%%%%%%%%%%%%%%%%%%%%%%%%%%%%%%%%%%%%%%%%%%%%%%%%%%%%%%%%%%%%%%%%%%%%%%%%%%%%%%%%%%%%%%%%%%%%%%
% ---- Figure--- Figure ---- Figure--- Figure ---- Figure--- Figure --- Figure--- Figure ----%
%%%%%%%%%%%%%%%%%%%%%%%%%%%%%%%%%%%%%%%%%%%%%%%%%%%%%%%%%%%%%%%%%%%%%%%%%%%%%%%%%%%%%%%%%%%%%%

%%%%%%%%%%%%%%%%%%%%%%%%%%%%%%%%%%%%%%%%%%%%%%%%%%%%%%%%%%%%%%%%%%%%%%%%%%%%%%%%%%%%%%%%%%%%%%
% ---- TABLE --- Table ---- Table--- Table ---- Table--- Table --- Table--- Figure ----%
%%%%%%%%%%%%%%%%%%%%%%%%%%%%%%%%%%%%%%%%%%%%%%%%%%%%%%%%%%%%%%%%%%%%%%%%%%%%%%%%%%%%%%%%%%%%%%

\newpage

\begin{table}[ht]
\centering
\caption{A summary of {\it XMM-Newton} (Xn, with n from 1 to 5; see Fig. \ref{fig:swiftlc}) and {\it Chandra} observations (C1) used in this paper.}
\begin{tabular}{||c c c c c||} 
 \hline
  & ObsID & Exposure (ks) & Date observed & N$_{\rm seg}^{\dagger}$ \\ [0.5ex] 
 \hline\hline
 & & & & \\
 X1 & 0694651201 & 23 & 2014-12-06 &  1 \\ 
 X2 & 0722480201 & 95 & 2014-12-08 & 1 \\
 X3 & 0694651401 & 25 & 2015-01-01 & 2 \\
 X4 & 0694651501 & 23.5 & 2015-07-10 & 0 \\
 X5 & 0770980101 & 96.5 & 2015-12-10 & 2 \\
  & & & & \\
 \hline 
 
  &  &  &  & \\
 C1 & 18345 & 25 & 2016-01-28 & 2 \\ [1ex] 
 \hline
\end{tabular}
\newline
$^{\dagger}$The number of uninterrupted 10 ks segments.
\label{table:table1}
\end{table}
 \vfill\eject
 
%%%%%%%%%%%%%%%%%%%%%%%%%%%%%%%%%%%%%%%%%%%%%%%%%%%%%%%%%%%%%%%%%%%%%%%%%%%%%%%%%%%%%%%%%%%%%%
% ---- TABLE --- Table ---- Table--- Table ---- Table--- Table --- Table--- Figure ----%
%%%%%%%%%%%%%%%%%%%%%%%%%%%%%%%%%%%%%%%%%%%%%%%%%%%%%%%%%%%%%%%%%%%%%%%%%%%%%%%%%%%%%%%%%%%%%%

\newpage
\begin{table}[ht]
\centering
\caption{Good time intervals that resulted from the criterion of selecting only data segments that were uninterrupted for over 10 ks (see Figs. \ref{fig:xmmlcs1} and \ref{fig:xmmlcs2}).}
\begin{tabular}{||c c c c||} 
 \hline
  & ObsID & start time (s)$^{\dagger}$ & end time (s)$^{\dagger}$ \\ [0.5ex] 
 \hline\hline
 & & & \\
 X1 & 0694651201 & 534306824 & 534316824  \\ 
 X2 & 0722480201 & 534445303 & 534455303  \\
 X3 & 0694651401 & 536535413 & 536545413 \\
 X3 & 0694651401 & 536545413 & 536555413 \\
 X5 & 0770980101 & 566193394 & 566203394   \\
 X5 & 0770980101 & 566203394 & 566213394 \\
  & & &  \\
 \hline 
 
  &  &  &  \\
 C1 & 18345 & 570380075 & 570390075 \\
 C1 & 18345 & 570390075 & 570400075 \\ [1ex] 
 \hline
\end{tabular}
\newline
$^{\dagger}$The start and stop times of the GTIs used in extracting the average PDS shown in Fig. \ref{fig:xmmchan_swift_pds}. These are measured in seconds since 1997-12-31T23:58:56.816 UTC , i.e., Modified Julian Date (MJD) of 50814.0, for both {\it XMM-Newton} and {\it Chandra} data. They have been rounded off to the nearest second. 
\label{table:table2}
\end{table}
 \vfill\eject
 
%%%%%%%%%%%%%%%%%%%%%%%%%%%%%%%%%%%%%%%%%%%%%%%%%%%%%%%%%%%%%%%%%%%%%%%%%%%%%%%%%%%%%%%%%%%%%%
% ---- TABLE --- Table ---- Table--- Table ---- Table--- Table --- Table--- Figure ----%
%%%%%%%%%%%%%%%%%%%%%%%%%%%%%%%%%%%%%%%%%%%%%%%%%%%%%%%%%%%%%%%%%%%%%%%%%%%%%%%%%%%%%%%%%%%%%%
\newpage

\begin{table}[ht]
\centering
\caption{Constraints on the red noise strength and QPO's statistical significance.}
\begin{tabular}{||c c c||} 
 \hline
  Red-noise slope & Normalization$^{\dagger}$ & Significance$^{*}$ \\ [0.5ex] 
 \hline\hline
 & &  \\
 -0.3 &  (1.9$\pm$1.9)$\times$10$^{-2}$&  $\approx$3.9$\sigma$ \\ 
 -0.7 &  (1.8$\pm$0.9)$\times$10$^{-3}$&  $>$4.1$\sigma^{\dagger\dagger}$ \\
 -1.0 &  (1.8$\pm$0.7)$\times$10$^{-4}$&  $>$4.1$\sigma^{\dagger\dagger}$ \\
 -1.4 &  (5.1$\pm$2.2)$\times$10$^{-6}$&  $>$4.1$\sigma^{\dagger\dagger}$ \\
 -1.7 &  (3.1$\pm$1.4)$\times$10$^{-7}$ &  $>$4.1$\sigma^{\dagger\dagger}$ \\ [1ex] 
 \hline
\end{tabular}
\newline
$^{\dagger}$The normalization was obtained by directly fitting ASASSN-14li's unbinned PDS with a model consisting of a power-law + constant + QPO (see sec. \ref{supsec:rednoise}). For the purposes of simulating the red noise curve and thus estimating the red-noise significance values, we used the upper limit on the normalization, i.e., best-fit normalization + 1$\sigma$ uncertainty. $^{*}$Global significance of the QPO in red noise power spectra simulated using the Monte Carlo methodology described in sec. \ref{supsec:rednoise} (also see Fig. \ref{fig:rednoisesigs}). $^{\dagger\dagger}$This is limited not by the strength of the QPO but by the number of simulations we could perform, which in our case was 50,000 (see Fig. \ref{fig:rednoisesigs}).

\label{table:powlawnoise}
\end{table}
 \vfill\eject
 
%%%%%%%%%%%%%%%%%%%%%%%%%%%%%%%%%%%%%%%%%%%%%%%%%%%%%%%%%%%%%%%%%%%%%%%%%%%%%%%%%%%%%%%%%%%%%%
% ---- TABLE --- Table ---- Table--- Table ---- Table--- Table --- Table--- Figure ----%
%%%%%%%%%%%%%%%%%%%%%%%%%%%%%%%%%%%%%%%%%%%%%%%%%%%%%%%%%%%%%%%%%%%%%%%%%%%%%%%%%%%%%%%%%%%%%%

\newpage
\begin{table}[t!]
\centering
\caption{Properties of the QPO at various epochs.}
\begin{tabular}{||c c c c c||} 
 \hline
  & ObsID & MJD$^{*}$ & rms$^{\dagger}$ & Fold-Period$^{\dagger\dagger}$\\ [0.5ex] 
 \hline\hline
 & & & & \\
 X1 & 0694651201 & 56998.11 & $>$1.3$\pm$0.3 & 127.7$\pm$0.5\\ 
 X2 & 0722480201 & 56999.77 & $>$1.4$\pm$0.3 & 136.8$\pm$0.3\\
 X3 & 0694651401 & 57024.02 & $>$1.1$\pm$0.3 & 133.3$\pm$0.1\\
 X4 & 0694651501 & 57213.40 & $>$3.3$\pm$0.8 & 123.4$\pm$0.3\\
 X5 & 0770980101 & 57367.28 & 2.8$\pm$0.7 & 136.7$\pm$0.1\\
 C1 & 18345 & 57415.737 & 35$\pm$8 &  134.6$\pm$0.1 \\
 \hline
\end{tabular}
\newline
$^{*}$Modified Julian Date. $^{\dagger}$The fractional rms amplitude of the QPO estimated from fitting a sinusoidal curve to the folded light curves. $^{\dagger\dagger}$Fold periods (in seconds) correspond to the frequency of the power spectral bin with the highest power within 7.65$\pm$0.7 mHz in the oversampled PDS (see sec. \ref{supsec:coherms}).
\label{table:tableqporms}
\end{table}
 \vfill\eject
\newpage

\newpage
\noindent\textbf{Supplementary Movie S1:} The movie shows gradual improvement in the QPO signal at 7.65 mHz as more data is added. This suggests that the QPO is present all throughout the outburst.

%-----
%

%\bibliography{supplement}
%
% -------

%%%%%%%%%%%%%%%%%%%%%%%%%%%%%%%%%%%%%%%%%%%%%%%%%%%
%%%%%%%%%%%%%%%%%%%%%%%%%%%%%%%%%%%%%%%%%%%%%%%%%%%
%%%%%%%%%%%%%%%%%%%%%%%%%%%%%%%%%%%%%%%%%%%%%%%%%%%

\end{document}